\documentclass{svjour3}                   
\smartqed  %
\usepackage{epsfig}
\usepackage{graphicx}
\usepackage{lineno}
\usepackage{amsmath}
\usepackage{hyperref}
\usepackage{setspace}
\usepackage{gensymb}
\usepackage[titletoc,toc,title]{appendix}
\usepackage[a4paper,width=130mm,top=25mm, bottom=25mm]{geometry}
\usepackage{color}
\journalname{{\rm submitted to} Wave Motion,}
\newcommand{\BM}[1]{\mbox{\boldmath $ #1 $}}
\newcommand{\myfrac}[2]{\displaystyle \frac{#1}{#2}}
\newcommand{\ii}{{\rm i}}

\def\RR{\vbox {\hbox to 8.9pt {I\hskip-2.1pt R\hfil}}}

\begin{document}

\title{On the viscoelastic-electromagnetic-gravitational analogy}

\author{Jos\'e M. Carcione$^{1,2}$ \and Jing Ba$^{1(*)}$}
\institute{
$^1$School of Earth Sciences and Engineering, 
Hohai University, Nanjing, 211100, China. \\
$^2$National Institute of Oceanography and Applied Geophysics - OGS, Trieste, Italy. \\
(*) Corresponding author
}

\date{Received: date / Accepted: date}

\maketitle

\baselineskip 23pt


\begin{abstract}
The analogy between electromagnetism and gravitation was achieved by linearizing the tensorial gravitational equations of general relativity and converting them into a vector form corresponding to Maxwell's electromagnetic equations. On this basis, we use the equivalence with viscoelasticity and propose a theory of gravitational waves. We add a damping term to the differential equations, which is equivalent to Ohm's law in electromagnetism and Maxwell's viscosity in viscoelasticity, to describe the attenuation of the waves. The differential equations in viscoelasticity are those of cross-plane shear waves, commonly referred to as SH waves.

A plane-wave analysis gives the phase velocity, the energy velocity, the quality factor and the attenuation factor of the field as well as the energy balance. To obtain these properties, we use the analogy with viscoelasticity; the properties of electromagnetic and gravitational waves are similar to those of shear waves. The presence of attenuation means that the transient field is generally a composition of inhomogeneous (non-uniform) plane waves, where the propagation and attenuation vectors do not point in the same direction and the phase velocity vector and the energy flux (energy velocity) are not collinear.
The polarization of cross-plane field is linear and perpendicular to the propagation-attenuation plane, while the polarization of the field within the plane is elliptical.

Transient wave fields in the space-time domain are analyzed with the Green function (in homogeneous media) and with a grid method (in heterogeneous media) based on the
Fourier pseudospectral method for calculating the spatial derivatives and a Runge-Kutta scheme of order 4 for the time stepping. In the examples, wave propagation at the Sun-Earth and Earth-Moon distances using quadrupole sources is considered in comparison to viscoelastic waves.
The Green and grid solutions are compared to test the latter algorithm.
Finally, an example of propagation in heterogeneous media is presented.
\end{abstract}


\section{Introduction}

The analogy between EM and acoustic-VE fields is well known (e.g. Carcione, 2022, Chapter 8). Carcione and Cavallini (1995a) have shown that the 2D Maxwell differential equations describing the propagation of the TM mode in anisotropic media are mathematically equivalent
to the SH-wave equation of an anisotropic solid with attenuation described by the Maxwell mechanical model. This equivalence was probably
known to Maxwell, who was aware of the analogy between the process of conduction
(static induction through dielectrics) and viscosity (elasticity). In geophysics, the analogy could be 
termed seismic-georadar analogy, since seismic (SH) and georadar waves are employed to 
evaluate the VE and EM properties of the near surface (Carcione and Cavallini, 1995b;  Carcione, 1996; Deidda and Balia, 2001; Daniels, 2004). 

Gravitational waves generated by accelerated masses distort space-time so that fixed distances change depending on the signal frequency. Einstein's general theory of relativity predicts that their velocity corresponds to the speed of light in a vacuum (Einstein and Rosen, 1937; Weber, 1961).
Maxwell (1865, Part IV, p. 492) attempted to use electromagnetic theory to understand the laws of gravity: ``After tracing to the action of the surrounding medium both the magnetic and
the electric attractions and repulsions, and finding them to depend on the inverse square
of the distance, we are naturally led to inquire whether the attraction of gravitation,
which follows the same law of the distance, is not also traceable to the action of a
surrounding medium." However, his approach leads to negative values for the energy.
He states: ``As energy is essentially positive, it is impossible for any part of space to have negative
intrinsic energy", ... ``As I am unable to understand in what way a medium can possess such properties, I cannot go any further in this direction in searching for the cause of gravitation".
Gravitational waves were first proposed by Heaviside (1893) as the equivalent of EM waves.
EM radiation generally has much smaller wavelengths than GR waves, whose frequency is in the ``audible' range (a few tens to a hundred Hz).

By studying Maxwell's equations, Lorentz was able to determine the form of the Lorentz transformations (Thid\'e, 2011, Section 7.1.1), which later formed the basis for Einstein's theory of gravitation.
Maxwell's (vector) equations are a good approximation for the description of GR waves, as they are compatible with the special theory of relativity (not with Galilean relativity). These equations therefore have the same form when transformed from one reference frame to another using the Lorentz transformation, while preserving the speed of light (which is the same for all observers).
In fact, Maxwell's equations can be transformed into a Lorentz covariance by arranging them in an antisymmetric -- Faraday -- tensor formulation (Thid\'e, 2011, eq. 7.78).

In principle, the physical analogy between the general theory of relativity and electromagnetism fails because there are two types of electric charges and only one type of mass, and because two particles with the same type of charge repel each other, while two particles with the same type of mass attract each other. Nevertheless,
GR waves arise from accelerated masses just as EM waves arise from accelerated charges, and a mathematical analogy can be made.
Forward (1961, Appendix), Peng (1983) and Wald (1984, Section 4.4) have shown that the linearized Einstein equations lead to the EM wave equation if one assumes that all velocities are small, so that special relativity can be neglected, and that all gravitational effects are weak (see also Carcione, 2022, Section 8.20). The GR speed is that of light (in empty space) or very close to it, as confirmed by the observation of the GR wave GW170817 (Abbott et al., 2017).
The electric field corresponds to the gravitational field, and the gravitational equivalent of the magnetic field has yet to be discovered experimentally. However, it is important to note that some linearized forms of the Lorentz force and potential energy differ by a factor of 4 from the analogous formula of general relativity, and in these cases there is no perfect isomorphism between the electromagnetic and gravitational equations with respect to the Lorentz force (Cattani, 1980, Eq. 45; Peng, 1983, Eq. 13).

L\'opez (2018, Eq. 10-13) developed an extended Newtonian theory of gravity analogous to Maxwell's equations, without referring to the general theory of relativity. He shows that several properties already known from electrodynamics (Poynting vector, energy density, tensor stress and radiation) are fully reproduced for GR fields.
On the other hand, Forward (1961) and Scharpf (2017), reporting on the analogy in Section 5.2.1 and equations 9.3.12 and 9.3.17, respectively, obtained tensorial gravitational equations.
Since there is no monopole or dipole contribution in gravitational radiation (lack of conservation of mass and momentum in contrast to the electromagnetic case), the first multipole moment that can be compared between the two types of radiation is the quadrupole (Price et al., 2013).
Price et al. (2013, Eq. 10) obtain a tensorial form of the gravitational equations in vacuum, where the tensors are traceless. They assume gravitostatics, i.e. the GR fields are weak and the masses (sources) do not move at velocities comparable to the speed of light (flat space-time).
In their formulation in spherical coordinates, if the tensor versions $E_{\phi \phi} - E_{\theta \theta}$ and
$B_{\phi \theta}$ are neglected, then the field equations for the gravitational scalars -- multiplied by $r$ -- are the same as those for electromagnetism.
Another tensor approach equivalent to Maxwell's equations is given in Barnett (2014, Eq. 29).
The radiation problem is discussed quantitatively in Prather (2020) and in Gomes and Rovelli (2024). Prather (2020, Eq. IV.32) obtains a negative Poynting vector (related to the negative energy) based on Heaviside's (1893) equations, indicating that the field gains energy instead of losing energy, as in the case of EM waves.

Here we consider the vector forms of the gravitational equations given by
Braginsky et al. (1977, eq. 3.8),
Peng (1983, eq. 10) and Ummarino and Gallerati (2019, eq. 23) from a linear approximation of general relativity.
In addition, we consider the wave attenuation (Ohm's law) according to Ciubotariu (1991). Alternative attenuation mechanisms are described by Baym et al. (2017), namely due to collisions in matter and Landau damping, where particles surf the GR wave and extract energy from it. Unlike solutions to the perturbed Einstein equations in vacuum, dispersive GR waves do not travel exactly at the speed of light. As a result, the GR wave can exchange energy with scalar massive particles in resonance (Asenjo and Mahajan, 2020). Weinberg (2004, Eq. 2) proposes a tensor wave equation that is similar to the scalar telegraph equation and contains a damping term due to the flow of free neutrinos.

He (2021) simulated GR waves propagating in a potential well in free space using a finite element method based on an acoustic-like wave equation.
Scharpf (2017) performs hybrid analytic-numerical simulations to model the time evolution of the source, which corresponds to the GR of binary black holes. Here we calculate the transient field using the Green function and grid methods to simulate seismic and georadar waves (e.g. Carcione, 2022, Chapter 9). The time history of the source is modeled phenomenologically with chirp functions.

\section{Space-time domain mathematical analogy} 

As already mentioned, Maxwell (1865) initially tried to develop a vector theory of gravity by analogy with his electromagnetic (EM) equations, but got nowhere because his static potential energy is negative, since two masses attract each other, unlike two electric charges of the same sign, which repel each other.
Heaviside (1893), pursued further Maxwell's attempt, and McDonald (1997)
indicates that Heaviside's gravity theory can be derived from general relativity. 
Here, we consider other vector forms of the gravitational equations. 

The analogy Maxwell made between electromagnetism and gravity implies that the GR-wave equations 
\begin{equation} \label{1}
\begin{array}{l}
\nabla \times {\bf g} = - b \partial_t {\bf h} , \\
\nabla \times {\bf h} = {\bf m} + d {\bf g} + a  \partial_t {\bf g} ,  \\
\nabla \cdot {\bf b} = \nabla \cdot (b{\bf h})= 0 , \ \ \ \nabla \cdot {\bf g} = a^{-1} \rho 
\end{array}
\end{equation}
(Braginsky et al., 1977, Eq. 3.8; 
Peng, 1983, Eq. 10; 
Behera and Naik, 2004, Eqs. 33-36; 
Hills, 2020, Eqs. 9-10)
are mathematically equivalent to the EM equations
\begin{equation} \label{2}
\begin{array}{l}
\nabla \times {\bf E} = - \mu \partial_t {\bf H} , \\
\nabla \times {\bf H} = {\bf J} + \sigma {\bf E} + \epsilon  \partial_t {\bf E} ,  \\
\nabla \cdot {\bf B} = \nabla \cdot (\mu {\bf H})= 0 , \ \ \ \nabla \cdot {\bf E} = \epsilon^{-1}  \rho_e 
\end{array}
\end{equation}
(Thid\'e, 2011, Eq. 1.48; Carcione, 2022; Eqs. 8.1-2), 
which differ from Heaviside (1893) equations by a sign in ${\bf h}$ (Behera and Barik, 2017, Eqs. 2 and 5), where the analogy and the symbols and their units in the SI system are given in Tables 1 and 2, respectively. We added a dissipation term $d {\bf g}$ in equation (\ref{1}) in analogy with  $\sigma {\bf E}$ (Ohm's law) according to Ciubotariu (1991). 
In the absence of damping (free space), the velocity is $c$ = 299792458 m/s $\approx$ 30 cm/ns, the light velocity. 

\section{Plane-wave analysis}

The plane-wave analysis gives the 
expressions of measurable quantities, such as the slowness vector, the 
energy-velocity vector and the quality factor as a function of frequency.  
Assume harmonic plane waves with a phase factor
\begin{equation} \label{p1}
\exp [ \ii \omega ( t -  {\bf s} \cdot {\bf x}  ) ] ,
\end{equation}
where $\omega = 2 \pi f$ is the angular frequency, $f$ is the frequency in Hz, ${\bf x}$ is the position vector, $\ii = \sqrt{-1}$, 
${\bf s} = {\bf k}/\omega$ is the complex slowness vector, 
\begin{equation}\label{p2}
{\bf k} = \omega {\bf s} = {\BM \kappa} - \ii {\BM \alpha} = \kappa \hat {\BM \kappa} - \ii \alpha \hat {\BM \alpha} 
\end{equation}
is the complex wavevector, with
${\BM \kappa}$ being the real wavevector and ${\BM \alpha}$ 
being the attenuation vector, which make an angle $\gamma$, the inhomogeneity angle. They express
the magnitudes of both the wavenumber $\kappa$ and the attenuation factor $\alpha$,
and the directions of the normals to planes of constant phase and planes of constant amplitude, respectively.
Waves for which the wavenumber and attenuation vectors do not point in the same direction are called inhomogeneous in viscoelasticity and non-uniform in 
electromagnetism. If the angle $\gamma$ between these vectors is zero, we have homogeneous or uniform plane waves for which 
\begin{equation}\label{p20}
{\bf k} = k \hat {\BM \kappa}, \ \ \ k = \kappa - \ii \alpha,
\end{equation}
where $k$ is the complex wavenumber. 

\subsection{Phase velocity, quality factor and attenuation factor}

We use the following correspondences between time and frequency domains:
\begin{equation} \label{p3}
\nabla \times \rightarrow - \ii \omega {\bf s} \times  \ \ \  {\rm and} \ \ \
\partial_t \rightarrow \ii \omega .
\end{equation}

For time-harmonic fields, 
equation (\ref{1}) reads
\begin{equation} \label{p4} 
\begin{array}{l}
\nabla \times {\bf g} = - \ii \omega {\bf b} , \\
\nabla \times {\bf g} = - \ii \omega b {\bf h} , 
\\ \\
\nabla \times {\bf h}  = \ii \omega {\bf w} + {\bf m}^\prime , \ \ \ {\bf w} \equiv a {\bf g} , \ \ \ {\bf m}^\prime \equiv  {\bf m} + d {\bf g} \\
\nabla \times {\bf h} = (\ii \omega a + d) {\bf g} + {\bf m} .
\end{array}
\end{equation}
Taking the vector product of equation (\ref{p4})$_4$ with ${\nabla}$ and 
use of (\ref{p4})$_2$ gives 
\begin{equation} \label{p5}
{\nabla} \times ( {\nabla} \times {\bf h} )  - \omega^2 \bar a b {\bf h} = 0, 
\end{equation}
where
\begin{equation} \label{p6}
\bar a = a - \frac{\ii d}{\omega}. 
\end{equation}
Since ${\bf \nabla} \times ( {\bf \nabla} \times {\bf h} ) = \nabla ( \nabla \cdot {\bf h} ) - \Delta {\bf h}$, and $\nabla \cdot {\bf h}$ = 0 in locally uniform media (see equation (\ref{1})$_3$, we have the wave equation 
\begin{equation} \label{p7}
\Delta {\bf h}  + k^2 {\bf h} = 0,  \ \ \ k = \frac{\omega}{v} , 
\end{equation}
where
\begin{equation} \label{p8}
v = \frac{1}{\sqrt{\bar a b}} 
\end{equation}
is the complex velocity. A similar wave equation can be obtained for ${\bf g}$ if $\nabla \cdot \rho$ = 0:
\begin{equation} \label{p5p}
{\nabla} \times ( {\nabla} \times {\bf g} )  - \omega^2 \bar a b {\bf g} = 0, 
\end{equation}

Substituting (\ref{p2}) into (\ref{p1}) gives 
\begin{equation} \label{p8p}
\exp ( - {\BM \alpha} \cdot {\bf x} ) \exp [ \ii \omega ( t -  {\BM \kappa} \cdot {\bf x}  ) ] .
\end{equation}
In analogy with viscoelasticity (Carcione, 2022, Eq. 3.35), we have 
\begin{equation}\label{p9}
\begin{array}{c}
2 \kappa^2 = {\rm Re} ( k^2 ) \left( 1 + \sqrt { 1 +  Q^{-2} {\rm sec}^2 \gamma } \right) , \\
2 \alpha^2 = {\rm Re} ( k^2 ) \left( -1 + \sqrt { 1 + Q^{-2} {\rm sec}^2 \gamma } \right) ,
\end{array}
\end{equation}
where
\begin{equation}\label{p10}
Q = - \frac{ {\rm Re} ( k^2 ) }{{\rm Im} ( k^2 ) } =  \frac{ {\rm Re} ( v^2 ) }{{\rm Im} ( v^2 ) } = 
\frac{ \omega a}{d} 
\end{equation}
is the quality factor for homogeneous waves ($\gamma$ = 0), where we have used equations (\ref{p6}) and (\ref{p8}). (Below, we will see that it also correspond to the quality factor of inhomogeneous waves). 
There is no damping if $d$ = 0, since $Q = \infty$. The phase velocity and attenuation factor are 
\begin{equation}\label{p11}
v_p = \frac{\omega}{\kappa} = \left[ {\rm Re} \left ( \frac{1}{v} \right) \right]^{-1} = \left[ {\rm Re}   ( \sqrt{\bar a b} ) \right]^{-1}
\end{equation}
and
\begin{equation}\label{p12}
\alpha = - \omega {\rm Im} \left ( \frac{1}{v} \right)  = - \omega {\rm Im} ( \sqrt{\bar a b} ) ,
\end{equation}
respectively, where the right-hand-sides hold for homogeneous waves, otherwise these quantities depend on $\gamma$ as in equation (\ref{p9}). 

The phase velocity in vector form is 
\begin{equation}\label{p111}
{\bf v}_p =   \frac{\omega {\BM \kappa}}{|{\BM \kappa}|^2} = v_p \hat {\BM \kappa} . 
\end{equation}

\subsection{Space-frequency-domain energy balance and time averages for inhomogeneous waves}

The scalar product of the complex conjugate of equation (\ref{p4})$_3$ with ${\bf g}$, 
use of $\nabla \cdot \ ( {\bf g} \times {\bf h}^\ast )$ = 
$( \nabla \times {\bf g} ) \cdot {\bf h}^\ast - {\bf g} \cdot ( \nabla \times 
{\bf h}^\ast )$, 
and substitution of equation (\ref{p4})$_1$ gives 
Umov-Poynting theorem for harmonic fields  
\begin{equation} \label{p13} 
- \nabla \cdot \ {\bf p} = 
\frac{1}{2} {{\bf m}^\prime}^\ast \cdot {\bf g} -
2 \ii \omega \left( \frac{1}{4} {\bf g} \cdot {\bf w}^\ast  
- \frac{1}{4} {\bf b} \cdot {\bf h}^\ast \right) , 
\end{equation}
where
\begin{equation} \label{p14}
{\bf p} = \frac{1}{2} {\bf g} \times {\bf h}^\ast
\end{equation}
is the complex Umov-Poynting vector. Substitution of the material properties into 
equation (\ref{p13}) yields
\begin{equation} \label{p15}
\nabla \cdot \ {\bf p} = 2 \ii \omega \left(
\frac{1}{4} \bar a^\ast |{\bf g}|^2  
- \frac{1}{4} b |{\bf h}|^2 \right) ,   
\end{equation} 
where we have assumed ${\bf m}$ = 0.
Each term has a precise physical meaning on a time-average basis: 
\begin{equation} \label{p16}
\begin{array}{l}
\myfrac{1}{4} {\rm Re} ( \bar a^\ast ) |{\bf g}|^2  =
\myfrac{1}{4} {\rm Re} ( \bar a ) |{\bf g}|^2  
\equiv \langle E_g \rangle  , \ \ \ \mbox{gravitational energy} \\ \\
\myfrac{\omega}{2} {\rm Im} ( \bar a^\ast ) |{\bf g}|^2  =
- \myfrac{\omega}{2} {\rm Im} ( \bar a) |{\bf g}|^2  
\equiv \langle \dot D_g \rangle  , \ \ \ \mbox{rate of dissipated gravitational energy} \\ \\
\myfrac{1}{4} b |{\bf h}|^2  \equiv \langle E_{cg} \rangle, \ \ \  \mbox{cogravitational energy}
\end{array}
\end{equation}
Substituting the preceding expressions into equation (\ref{p15}), yields
the energy-balance equation
\begin{equation} \label{ebeq}
\nabla \cdot \ {\bf p} - 2 \ii \omega ( \langle E_g \rangle - \langle E_{cg} \rangle ) + 
\langle \dot D_g \rangle  = 0 .
\end{equation} 
The quality factor is obtained as twice the stored (strain) energy $\langle E_g \rangle$ divided by the dissipated energy $\langle D_g \rangle = \langle \dot D_g \rangle / \omega$ in analogy with viscoelasticity and electromagnetism (Carcione, 2022; Eqs. 3.125 and 8.341). This gives equation (\ref{p10}). The Umov-Poynting theorem provides a consistent formulation of energy flow.
In the following, we present more explicit expressions of the energy balance in the wavenumber (slowness)-frequency domain. 

\subsection{Space-time domain energy balance}

Next, we exploit the analogy between VE and EM waves. 
Let us consider the 
$(x,z)$-plane. 
If the medium properties are constant, the EM equations have two decoupled solutions, namely 
$ E_1 $, 
$ E_3 $ and 
$ H_2 $ are               
decoupled from                                                                  
$ E_2 $, 
$ H_1 $ and 
$ H_3 $, 
where the subindices 1, 2 and 3 are equivalent to $x$, $y$ and $z$, respectively.
The first three fields obey the TM (transverse-magnetic) differential equations, while the second ones obey the TE (transverse-electric) equations (e.g. Carcione, 2022, Fig. 8.11). Without loss in generality it is enough to consider one of these differential equations. 
From the acoustic-electromagnetic analogy (Carcione and Cavallini, 1995a; Carcione, 2022, Chapter 8) it is well known that  TM (transverse-magnetic) waves 
are mathematically equivalent to VE type-II S (shear) waves, The complete analogy is
\begin{equation} \label{p22}
\begin{array}{ll}       
\partial_1 \sigma_{12} - \partial_3 (-\sigma_{23})  = \rho \partial_t v , & {\rm VE} \\                                                             
\partial_1 E_3 - \partial_3 E_1 = \mu \partial_t H , & {\rm EM} \\    
\partial_1 g_3 - \partial_3 g_1 = b \partial_t h , & {\rm GR} \\  
\\                          
- \partial_3 v = \eta^{-1} (- \sigma_{23} )  + R^{-1} \partial_t (- \sigma_{23} ) , & {\rm VE}  \\     
- \partial_3 H = \sigma E_1 + \epsilon \partial_t E_1 , &{\rm EM}\\      
- \partial_3 h = d g_1 + a \partial_t g_1 ,  &{\rm GR} \\      
\\                   
\partial_1 v = \eta^{-1} \sigma_{12}  + R^{-1} \partial_t \sigma_{12} ,  & {\rm VE}\\
\partial_1 H = \sigma E_3 +  \epsilon \partial_t E_3 , &{\rm EM}\\
\partial_1 h = d g_3 +  a \partial_t g_3 , & {\rm GR}, 
\end{array}
\end{equation}
where $H = H_2$, $v$ and $h = h_2$ are the anti-plane magnetic, particle-velocity and cogravitational components, i.e., those components perpendicular to the propagation $(x,z)$-plane, $\sigma_{ij}$ are off-diagonal stress components, 
$R$ is the shear rigidity or shear modulus, and $\eta$ is the Maxwell viscosity. It is clear that the equivalence are $H \Leftrightarrow v$, $E_1 \Leftrightarrow - \sigma_{23}$, $E_3 \Leftrightarrow \sigma_{12}$, $\epsilon \Leftrightarrow R^{-1}$, $\sigma 
\Leftrightarrow \eta^{-1}$ and $\mu \Leftrightarrow \rho$. 

Based on the VE-EM-GR analogy (see Tables 1 and 2), the energy balances in the space-time domain are
\begin{equation}\label{ebal}
\begin{array}{cll}     
- \nabla \cdot \ {\bf p}
 & = \partial_t ( T  + V ) + \dot D , & {\rm VE} \\ 
 & = \partial_t ( E_m  + E_e ) + \dot D_c,  & {\rm EM} \\
& = \partial_t ( E_{cg}  + E_g ) + \dot D_g,  & {\rm GR}, 
\end{array}     
\end{equation}
(Carcione, 2022, Eqs. 3.75 and 8.99), where $T$ is the kinetic energy, $V$ is the strain energy, $\dot D$ is the rate of dissipated viscoelastic  energy, 
$E_m$ is the stored magnetic energy, $E_e$ is the stored electric energy,  
$D_c$ is the dissipative conductive energy, 
$E_{cg}$ is the stored cogravitational energy, $E_g$ is the stored gravitational energy  
and $D_g$ is the dissipative gravitational energy. 

The complex shear modulus $\bar R$ associated with equations (\ref{p22}) is given by the Maxwell mechanical model (Carcione, 2022, Eq. 2.167), i.e., 
\begin{equation} \label{p23}
\frac{1}{\bar R} = \frac{1}{R} - \frac{\ii}{\omega \eta} \Leftrightarrow 
\bar \epsilon = \epsilon - \frac{\ii \sigma}{\omega}
\Leftrightarrow 
\bar a = a - \frac{\ii d}{\omega},
\end{equation}
while 
\begin{equation} \label{p23p}
\rho \Leftrightarrow \mu \Leftrightarrow b . 
\end{equation}
Then, in analogy with VE waves, the polarization of the cross-plane EM and GR waves is linear perpendicular to the $({\BM \kappa}, {\BM \alpha})$-plane as illustrated in Fig. 3.3. of Carcione (2022), and for inhomogeneous plane waves:  
\begin{align} 
\langle {\bf p} \rangle & =  \frac{1}{2} \omega |\phi_0|^2  \exp ( -
2 {\BM \alpha} \cdot {\bf x} ) [ b \omega^2 {\BM \kappa}  + 2 (
{\BM \kappa} \times  {\BM \alpha} ) \times (\bar R_I{\BM \kappa}
- \bar R_R{\BM \alpha} ) ] , \notag \\ 
\langle T \rangle \Leftrightarrow
\langle E_m \rangle \Leftrightarrow
\langle E_{cg} \rangle  & =
\frac{1}{4} b \omega^2 |\phi_0|^2 \exp ( - 2 {\BM \alpha} \cdot {\bf x} )(|{\BM \kappa}|^2 + |{\BM \alpha}|^2 ) , \notag \\
\langle V \rangle \Leftrightarrow
\langle E_e \rangle \Leftrightarrow
\langle E_g \rangle & = \frac{1}{4}|\phi_0|^2  \exp ( - 2 {\BM \alpha} \cdot {\bf x} )
[ b \omega^2 (|{\BM \kappa}|^2 - |{\BM \alpha}|^2 ) + 4 \bar R_R |{\BM \kappa} \times {\BM \alpha}|^2  ] , \label{genP} \\
\langle E \rangle & = 
\langle E_g \rangle + \langle E_{cg} \rangle  =
 \frac{1}{2}|\phi_0|^2  \exp ( - 2 {\BM \alpha} \cdot {\bf x} )
[ b \omega^2 |{\BM \kappa}|^2 + 2 \bar R_R |{\BM \kappa} \times {\BM \alpha}|^2  ] , \notag \\
\langle \dot D \rangle \Leftrightarrow
\langle \dot D_c \rangle \Leftrightarrow
\langle \dot D_g \rangle & = \omega |\phi_0|^2  \exp ( - 2 {\BM \alpha} \cdot {\bf x} )
[ b \omega^2 ({\BM \kappa} \cdot {\BM \alpha} ) + 2 \bar R_I |{\BM \kappa} \times {\BM \alpha}|^2  ] \notag = \omega \langle D_g \rangle \\
\langle D_g \rangle & = |\phi_0|^2  \exp ( - 2 {\BM \alpha} \cdot {\bf x} )
[ b \omega^2 ({\BM \kappa} \cdot {\BM \alpha} ) + 2 \bar R_I |{\BM \kappa} \times {\BM \alpha}|^2  ] \notag 
\end{align}
(Carcione, 2022, Eqs. 3.132; correcting a typo in these equations, it should be 
$\Xi_0 = \Gamma_0/ \sqrt{ {\bf k} \cdot {\bf k}^\ast }$),
where $\phi_0$ is a constant amplitude, with units [m$^2$] in viscoelasticty, [A s] in electromagnetism, and
[kg] in gravity, 
\begin{equation} \label{p24}
\bar R_R = 
\frac{(\omega \eta)^2 R}{R^2 + (\omega \eta)^2} \Leftrightarrow
\frac{\epsilon}{\epsilon^2 + (\sigma/\omega)^2} \Leftrightarrow
\frac{a}{a^2 + (d/\omega)^2} 
\end{equation}
and 
\begin{equation} \label{p24p}
\bar R_I = 
\frac{\omega \eta R^2}{R^2 + (\omega \eta)^2} \Leftrightarrow
\frac{\sigma/\omega}{\epsilon^2 + (\sigma/\omega)^2} \Leftrightarrow
\frac{d/\omega}{a^2 + (d/\omega)^2} 
\end{equation}
in terms of the VE-EM-GR properties, where the subindices $R$ and $I$ denote real and imaginary parts. The energies $E_g$ and $E_{cg}$ are the equivalent of the strain and kinetic energies in viscoelasticity. 

The energy-velocity vector, ${\bf v}_e$, is given by the energy power flow, 
${\rm Re} ({\bf p})$, divided by the total stored energy density,
\begin{equation} \label{p24}
{\bf v}_e = \frac{{\rm Re}({\bf p})}{ \langle E_g  + E_{cg} \rangle} = 
\frac{\langle {\bf p} \rangle}{ \langle E \rangle}. 
\end{equation}

Substitution of the corresponding expressions yields 
\begin{equation} \label{p25}
{\bf v}_e = 
\frac{b \omega^3 {\BM \kappa}  + 2 \omega (
{\BM \kappa} \times  {\BM \alpha} ) \times (\bar R_I{\BM \kappa}
- \bar R_R{\BM \alpha} )}
{b \omega^2 |{\BM \kappa}|^2 + 2 \bar R_R |{\BM \kappa} \times {\BM \alpha}|^2} .
\end{equation}
Comparison of this expression for the energy velocity with the corresponding
expressions for phase velocity indicated by (\ref{p111}) shows that the energy velocity
is not equal to the phase velocity in either direction or amplitude for inhomogeneous waves.
For homogeneous waves ${\BM \kappa} \times {\BM \alpha}$ = 0 and ${\bf v}_e = {\bf v}_p$ (see equation (\ref{p111})).  
Relation $\hat {\BM \kappa} \cdot {\bf v}_e = v_p $ holds in general (Buchen, 1971, Eq. 40; Krebes and Le, 1994, Eq. B23; Carcione, 2022, Eq. 3.123)
and other similar relations (Carcione, 2022, Eqs. 3.117-3.121).

On the other hand, the quality factor is given by 
\begin{equation} \label{q1}
Q = \frac{2 E_g}{ \langle D_g \rangle} = \frac{1}{2} \cdot
 \frac{b \omega^2 (|{\BM \kappa}|^2 - |{\BM \alpha}|^2 ) + 4 \bar R_R |{\BM \kappa} \times {\BM \alpha}|^2}{ b \omega^2 ({\BM \kappa} \cdot {\BM \alpha} ) + 2 \bar R_I |{\BM \kappa} \times {\BM \alpha}|^2 } 
\end{equation}

Let us consider the propagation and attenuation vectors in a Cartesian system, 
\begin{equation} \label{E1}
\begin{array}{l}
{\BM \kappa} = \kappa (l_x , 0, l_z )^\top \equiv \kappa  (\sin \theta, 0, \cos \theta)^\top,  \\
{\BM \alpha} = \alpha (m_x , 0, m_z )^\top \equiv \alpha   [\sin (\theta + \gamma), 0, \cos (\theta + \gamma) ]^\top, 
\end{array}
\end{equation}
where $\theta$ is the propagation angle, $\gamma$ is the inhomogeneity angle and $\kappa$ and $\alpha$ must be obtained from equations (\ref{p9}). Then
\begin{equation} \label{E2}
\begin{array}{l}
{\BM \kappa} \times {\BM \alpha} = (\kappa_z \alpha_x - \kappa_x \alpha_z) \ \hat {\bf y} = \kappa \alpha \sin \gamma \ \hat {\bf y} , \\
({\BM \kappa} \times {\BM \alpha}) \times {\BM \kappa} = \kappa^2 \alpha \sin \gamma 
(l_z, 0, - l_x)^\top, \\
({\BM \kappa} \times {\BM \alpha}) \times {\BM \alpha} = \kappa \alpha^2 \sin \gamma 
[m_z, 0, - m_x)]^\top, 
\end{array}
\end{equation}
Substituting equations (\ref{E2}) into equation (\ref{p25}), we obtain
\begin{equation} \label{E3}
\begin{array}{l}
{\bf v}_e = 
v_{ex} \hat {\bf x} + v_{ez} \hat {\bf z} , \\ \\
v_{ex} = \myfrac{\omega}{\kappa} \cdot \myfrac{[ b l_x \omega^2 +2  \alpha \sin \gamma (\kappa l_z \bar R_I - \alpha m_z \bar R_R) ]} {b \omega^2  + 2  \alpha^2 \bar R_R \sin^2 \gamma} , \\ \\
v_{ez} = \myfrac{\omega}{\kappa} \cdot \myfrac{[ b l_z \omega^2 +2  \alpha \sin \gamma (-\kappa l_x \bar R_I + \alpha m_x \bar R_R) ]} {b \omega^2  + 2  \alpha^2 \bar R_R \sin^2 \gamma} .
\end{array}
\end{equation}
For homogeneous waves, $\gamma$ = 0, and $v_e = \omega / \kappa = v_p$, the phase velocity, since $l_x^2 + l_z^2$ = 1. 

The quality factor (\ref{q1}) is given by 
\begin{equation} \label{q2}
Q  = \frac{1}{2} \cdot
 \frac{b \omega^2 (\kappa^2 - \alpha^2 ) + 4 \bar R_R \kappa^2 \alpha^2 \sin^2 \gamma}{ b \omega^2 \kappa \alpha \cos \gamma  + 2 \bar R_I \kappa^2 \alpha^2 \sin^2 \gamma} 
\end{equation}
This expression has been shown to be independent of $\gamma$ in Carcione et al. (2020) (see Carcione, 2022, Eq. 3.135) and equal to the quality factor of homogeneous waves  (\ref{p10}). 

\subsection{Polarization}

In general relativity, GR waves are transverse and have two linear polarizations (Thorne, 1994; Barnett, 2014) that can be combined to produce an elliptical polarization. In a binary star system, the amplitude perpendicular to the orbital plane (``face-on") is twice as large as the amplitude ``edge-on". The face-on radiation consists of equal amounts of the two gravitational wave polarizations, but they are out of phase to produce a circularly polarized wave. The edge-on wave contains only the linear polarization state and therefore has half the amplitude.
In the case of two black holes or neutron stars, for example, the plane of the orbital system is inclined at an angle between 0 and $\pi$/2 to the observer's line of sight. The first case corresponds to a system which is face-on to the observer, and the other to a system which is edge-on to the observer. In the latter case, only the linear polarization can be detected.
Circular polarization results from face-on radiation and equal amplitude (and out of phase) of the two linear polarizations, while  
a binary system seen at an angle with masses seen going around an ellipse generally gives an elliptical polarization. 

Let us consider the present theory and the analogy (\ref{p22}) with VE and TM waves.
The polarization of the GR waves ${\bf h}$ is linear and perpendicular to the $({\BM \kappa}, {\BM \alpha})$-plane (lying in the $(x,z)$-plane), as shown in
Figure 1a (see also Fig. 3.3 in Carcione, 2022). But what is the polarization of vector ${\bf g}$? 
Since ${\bf h} = h \hat {\bf y}$, using equation (\ref{p4})$_4$ (without source), we have
\begin{equation} \label{pol1}
{\bf g} = (\ii \omega \bar a)^{-1} \nabla \times {\bf h} = - (\omega \bar a)^{-1} {\bf k} \times {\bf h} = - h (\omega \bar a)^{-1} ( {\BM \kappa} \times \hat {\bf y} - \ii {\BM \alpha} \times \hat {\bf y} ) .
 \end{equation}
Vectors ${\BM \kappa} \times \hat {\bf y}$ and ${\BM \alpha} \times \hat {\bf y}$ lie in the $(x,z)$-plane and make an angle 
$\pi/2$ with ${\BM \kappa}$ and ${\BM \alpha}$, respectively. Let us assume that 
${\BM \kappa}$ is collinear with $\hat z$ and that ${\BM \alpha}$ lies in the positive $(x,z)$ quadrant. Then, 
${\BM \kappa} \times \hat {\bf y}$ is collinear with $- \hat {\bf x}$, and 
\begin{equation} \label{pol1}
{\bf g} =  h (\omega \bar a)^{-1} [ \kappa \hat {\bf x} + \ii \alpha ( - \hat {\bf x} \cos \gamma + \hat {\bf z} \sin \gamma ) ]
= 
 h (\omega \bar a)^{-1} [ ( \kappa  - \ii \alpha \cos \gamma) \hat {\bf x}+ \ii \alpha \sin \gamma \ \hat {\bf z} ]
 \end{equation}
or 
\begin{equation} \label{pol2}
\bar {\bf g} \equiv h^{-1} {\bf g} =
\left( \frac{\kappa  - \ii \alpha \cos \gamma}{\omega a - \ii d} \right)  \hat {\bf x}+ \left( \frac{\ii \alpha \sin \gamma}
{\omega a - \ii d} \right) \ \hat {\bf z}  \equiv X \hat {\bf x} + Z \hat {\bf z}.
 \end{equation}
Then, 
\begin{equation} \label{pol3}
\begin{array}{l}
\bar {\bf g} =|X| \exp (\ii \theta_1 ) \hat {\bf x} + |Z| \exp (\ii \theta_3 ) \hat {\bf z} , \\
\ \ \ \ \ \ \theta_1 = {\arctan} (X_I/X_R ) , \ \ \theta_3 = {\arctan} (Z_I/Z_R ) , 
\end{array}
\end{equation}
or, in the form of Jones vectors
\begin{equation} \label{pol4}
\begin{array}{l}
\bar {\bf g} =g_{\rm eff} [ A \hat {\bf x} + B \exp (\ii \delta) \hat {\bf z} ] , \\
g_{\rm eff} = \sqrt{|X|^2+|Z|^2} \exp (\ii \theta_1 ), \\
A=\myfrac{|X|}{\sqrt{|X|^2+|Z|^2}} , \ \ \
B=\myfrac{|Z|}{\sqrt{|X|^2+|Z|^2}} , \\
\delta = \theta_3 - \theta_1 
\end{array}
\end{equation}
(e.g. Peatross and Ware, 2024, Section 6.2). 
This implies that ${\bf g}$ is generally elliptically polarized in the $(x,z)$-plane, with semiaxes
\begin{equation} \label{pol5}
\begin{array}{l}
g_\psi = |g_{\rm eff}| \sqrt{A^2 \cos^2 \psi + B^2 \sin^2 \psi + A B \cos \delta \sin 2 \psi} , \\
g_{\psi \pm \pi/2} = |g_{\rm eff}| \sqrt{A^2 \sin^2 \psi + B^2 \cos^2 \psi + A B \cos \delta \sin 2 \psi} , \\
\psi = \frac{1}{2} {\arctan} \left( \myfrac{2 A B \cos \delta}{A^2-B^2}\right) 
\end{array}
\end{equation}
(Peatross and Ware, 2024, Section 6.3). 

The polarization is characterized by the ellipticity, given by the
ratio of the minor axis to the major axis:
\begin{equation}\label{ellip}
e = \frac{g_{\rm min}}{g_{\rm max}} .
\end{equation}
Figure 1b shows the ellipticity of the vector ${\bf g}$ as a function of $\gamma$,
it lies between zero for $\gamma$ = 0 (which corresponds to a linear polarization)
and approaches one for $\gamma$ = $\pi$/2 (corresponding to circular polarization).
We have assumed $Q$ = 10 ($d = \omega a/Q$) and $f$ = 100 Hz.

By analogy, the electric field ${\bf E}$ is elliptically polarized and all equations also apply to the EM case. If $\gamma$ = 0 (homogeneous wave), then ${\bf g}$ is linearly polarized along the $x$-direction.

\subsection{Computation of transient fields} 

Transient wave fields in the space-time domain are calculated using the Green function (homogeneous media) on the one hand, and a grid (mesh) method based on the Fourier pseudospectral method for calculating the spatial derivatives and a Runge-Kutta 4th-order scheme for heterogeneous media on the other. The Green function method uses an FFT subroutine from the NASA LaRC Computer Manual, Vol. II Section E2.4 (1975), and requires a complex vector of length a power of two.
The second (mesh) algorithm is explained in Sections 9.2.3. and 9.3.2. of Carcione (2022).

\section{Examples}

\subsection{Velocity of plane waves}

Let us first assume EM waves, where the instrument is a georadar with frequency $f$ = 500 MHz and a wet clay medium, for which $\epsilon = 10 \ \epsilon_0$, $\mu = \mu_0$ and $Q$ = 1, giving $\sigma = \omega \epsilon/Q = 2 \pi f \epsilon/Q$ = 0.11 S/m. Figure 2 shows the velocities as a function of the inhomogeneity angle $\gamma$. The velocity for homogeneous waves is that at $\gamma$ = 0. As can be seen, 
the field is strongly inhomogeneous. We do not have values for $d$ (the gravitational damping), but the physics should be similar, at least qualitatively.  

\subsection{Transient fields} 

Next, we perform simulations of VE and GR waves. 
The Einstein Toolkit \url{https://einsteintoolkit.org}
is a community-driven software platform to simulate gravitational waves among other things. Moreover, phenomenological signals parameterized by two physical parameters of a binary system are given in Ajith et al. (2008), namely, the total mass $M$ and the symmetric mass ratio $M_1 M_2/M^2$, where the signals are defined in the frequency domain.

The inspiral of the coalescence of two black holes (BHs) has the lowest frequencies. The amplitude and frequency of the wave increase continuously during the inspiral (``chirp" signal), together with the speed of the two BHs.
After the inspiral phase, the BHs begin to merge, the amplitude has its maximum at the merger and the frequency increases. Strictly, numerical relativity simulations are required to model this nonlinear process.
After the two BHs have merged, the signal is attenuated by the emission of energy in the form of gravitational waves

\subsubsection{Source time history}

A simpler phenomenological time history of the source can be obtained with chirp signals as indicated in appendix A (equation (\ref{A8})). Let us assume $\phi_0$ = 0, $t_0$ = 1 ns, $t_1$ = 0.15 s,  $t_2$ = 0.1 s, $f_0$ = 40 Hz, $f_1$ = 200 Hz, $a_1$ = 5/s, and $a_2$ = $-$100/s. Figure 3 shows the source time histories corresponding to linear, exponential and hyperbolic chirps (a) and the normalized spectrum (square of the absolute value of the Fourier transform) for the linear chirp (b), obtained with a signal duration of 0.6 s. The solid blue line corresponds to $a_1 = a_2$ = 0. 
The signal in (a) resembles that of a coalescence of two non-spinning black holes, as reported in Scharpf (2017, Fig. 7.4b)
The Fresnel ripples on the chirp spectrum (see blue line) arise because of the sudden discontinuities in the waveform at the beginning and end of the pulse. They are more in evidence for low values of the product $(t_1 - t_0) (f_1-f_0)$. In (b), we have considered $t_1$ = 0.6 s, such that the product is 96, but constant $c$ in equation (\ref{A2}) has been computed with $t_1$ = 0.15 s.  

\subsubsection{Green-function solution}

Figure 4 shows the solution using the Green function (see Appendix B) with $d$ = 0 (no loss) and an isotropic (point) source. The source-receiver distance is 
$r$ = 149597870700 m (the Sun-Earth distance, approximately 150 million km). Light velocity is 
$c$ = 2.99792458 $\times$ 10$^8$ m/s (nearly 30 cm/ns). The signal employs 8.32 minutes or 499 s to travel that distance, which is the time the Earth would remain in its elliptical orbit before flying off in a straight line, if the Sun were spontaneously removed from existence.
The length of the discrete fast Fourier transform is 2$^{20}$ = 1048576 and the sampling rate is $dt$ = 6 $\times$ 10$^{-4}$ s.  The field is normalized. 

Let us consider field attenuation. According to equation (\ref{p10}), the loss factor is 
\begin{equation}\label{s1}
d = \frac{ 2 \pi \bar f a}{Q} ,
\end{equation}
where $Q$ is the quality factor. The loss implies velocity dispersion, i.e., frequency dependence. To illustrate the physics, we consider a very low $Q$ = 10 and $\bar f$ = 120 Hz in equation (\ref{s1}). Figure 5 shows the phase velocity and attenuation factor as a function of frequency, where we can see that 
the velocity is zero at zero frequency (if $\omega \rightarrow$ 0, $\bar a \rightarrow - \ii \infty$, $v \rightarrow$ 0, and $v_p \rightarrow$ 0; see equations (\ref{p6}), (\ref{p8}) and (\ref{p11})) and approaches the velocity of light (30 cm/ns) at high frequencies.
To compute the solution based on the Green function, we consider a more realistic value, namely, $Q$ = 3 $\times$ 10$^5$.
Figure 6 shows the normalized signal displayed in Figure 4 (solid line) and the signal with attenuation (dashed red line).

\subsubsection{Grid-method solution}

We first simulate seismic waves in a medium with shear-wave velocity of $c$ = 3000 m/s, a density of $\rho$ = 2000 kg/m$^3$
and a quality factor $Q$ = 80, such that the Maxwell viscosity is
\begin{equation}\label{gs1}
\eta = \frac{R Q}{2 \pi \bar f} = \frac{\rho c^2 Q}{2 \pi \bar f}
\end{equation}
[see equation (\ref{p10}) and Carcione (2022, Eq. 2.176)], where $\bar f$ = 80 Hz. 
According to the Nyquist theorem, the maximum grid spacing must fulfil the following condition
\begin{equation}\label{gs3}
dx_{\rm max} = \frac{c_{\rm min}}{2 f_{\rm max}}
\end{equation}
(e.g. Carcione, 2022, Section 9.7), where $c_{\rm min}$ is the minimum velocity in the grid (two points per wavelength) and $f_{\rm max}$ is the maximum frequency.
If we consider 3000 m/s and 200 Hz, we obtain $dx \le$ 7.5 m.
The mesh has 442 $\times$ 442 grid points and the grid spacing is $dx$ = $dz$ = 7 m. 
The source is a quadrupole force in equation (\ref{p22})$_1$ ($\rho \partial_t v + {\rm force}$) with time history as in Figure 3 (solid line), located at the center of the mesh. Basically, grid points (221,220) and (221,222) have negative signs and grid points (220,221) and (222,221) have positive signs.
The time step is first chosen according to the stability condition and then according to the accuracy, so that 
\begin{equation}\label{gs2}
dt_{\rm max} = \frac{2 dx_{\rm min}}{\pi  c_{\rm max}}
\end{equation}
(Carcione, 2022, Eq. 9.12). In this case, $dt_{\rm max}$ = 1.15 ms. We consider $dt$ = 0.5 ms and a maximum time of 0.49 s.
Figure 7 shows snapshots of the component $v$ in the lossless (a) and lossy (b) cases, where we observe that the field has been attenuated in the latter case. 

We test the modeling code simulating gravitational waves and comparing the field component $h$ with the Green-function solution in homogeneous media.
The source is a point force with the time history of the preceding simulation. 
According to equation (\ref{gs3}), if we consider the speed of light and 200 Hz, we have $dx \le$ 750 km. Let us assume propagation distances comparable to those between the Earth and the Moon: 384400 km (the wave spends 1.28 s from the Earth to the Moon).
A mesh with 1105 $\times$ 1105 grid points and $dx$ = $dz$ = 700 km covers this distance. We have $a$ = 1.2 $\times$ 10$^{9}$ s$^2$ kg/m$^3$ and $b$ = 9.3 $\times$ 10$^{-27}$ m/kg and we assume $Q$ = 150 so that the loss factor is given in equation (\ref{s1}) with $f = \bar f$ = 80 Hz, specifically $d$ = 4 $\times$ 10$^9$ s kg/m$^3$. 
We consider $dt$ = 0.2 ms. 
The source and detector are located at grid points (552,552) and (709,709), respectively, so that the source-receiver distance is 155422.5 km (or 0.52 light-seconds). 
Figure 8 shows the comparison between the grid and Green solutions, where we can see that the agreement is excellent.

Let us consider the simulation of gravitational waves and a
quadrupole source as in Price et al. (2013) and Scharpf (2017). We assume $Q = \infty$ ($d$ = 0) in the upper half space and $Q$ = 150 in the lower half space, so that the loss factor is $d$ = 4 $\times$ 10$^9$ s kg/m$^3$. Figure 9 shows a snapshot of the component $h$ at 1.2 s, where we can see the attenuation below the interface. This problem has no analytical solution and the only solver is numerical or semi-analytical if reflectivity methods are used (e.g. M\"uller, 1985), but the latter is valid only for flat and parallel interfaces. On the other hand, grid methods can handle arbitrary interface geometries, as shown in the next example.

The general theory of relativity predicts that GR waves propagate in a vacuum at the speed of light. However, in media other than a vacuum, this speed should be affected by inhomogeneities (e.g. Pandey et al., 2022), but gravity is a curvature of spacetime caused by mass, not a force (like sound) traveling through space. Therefore, it is inappropriate to compare their propagation directly with that of light or sound.
However, GR waves, like EM waves, can be focused by gravitational lensing, and by analogy we can imagine that mass produces an ``effective index of refraction" due to the curvature of spacetime. The analogy with EM waves would mean that GR waves are ``slowed down" in a medium in the same way as EM waves. In this interpretation, we can vary the velocity $c$ in the constant $b$ (the gravitational constant $G$ is constant throughout space, so $a$ cannot vary). As in the previous example, the damping $d$ can also be assumed to be spatially variable, but in the next example we consider $d$ = 0.

Figures 10 and 11 show a snapshot at 0.9 s (a) and time histories (b) of the field components due to a point source. The velocity inside the circle is 0.8 $c$ (outside is $c$) and the source (white S) and receiver (white R) are indicated in the snapshot. At 0.9 s, the field has not yet reached the receiver, but one can see a reflection event from the circle (indicated by the black R). The field $h$ and $g_1$ are in phase, while $g_3$ is out of phase with the other components. After the main event in (a), one can observe reverberations (or multiples) arising inside the circle.

\subsection{Memory limitations of the proposed methods}

There is a memory limit for computers to propagate the signal distances of parsecs
(1 pc (parsec) = 3.26 light years = 3.0857 $\times$ 10$^{16}$ m), with the proposed algorithms. 
In the lossless case in homogeneous media, this is not a problem, as we can obtain the solution by a temporal convolution (\ref{B5}). In the lossy case, the Green-function solution in homogeneous media must be calculated in the frequency domain using the FFT. The propagation of signals with a maximum frequency of $f_{\rm max}$ = 200 Hz requires sampling rates of $dt$ = 1/(2$f_{\rm max}$) = 0.0025 s to avoid aliasing problems. 

On February 11, 2016, the scientific collaboration of LIGO and Virgo announced the first direct detection of gravitational waves from a binary star system (event GW150914, see e.g. Carcione (2022, Section 8.20)) with an average distance of $r$ = 400 Mpc =
12 $\times$ 10$^{24}$ m. The maximum time required for the signal to reach the Earth is $t_{\rm max} = r/c \approx$ 4 $\times$ 10$^{16}$ s, which divided by $dt$ gives 1.6 $\times$ 10$^{19}$ sampling points or about 2$^{64}$.
In a code, a dimension of real numbers takes 8 bytes, so that the number of bytes needed to define a single  array for the FFT is 8 $\times$ 2$^{64}$ =  2$^{67}$ $\approx$ 10$^{20}$, which equals to 10$^{5}$ PB (petabytes) = 10$^{11}$ MB (megabytes). The world's fastest computer has nearly 10 PB of RAM memory.

When calculating fields with the Green function (homogeneous media), however, these limitations are less problematic, as methods and approximations can be used to overcome this problem. On the other hand, wave propagation in heterogeneous media (e.g. if there are  galaxies, interstellar clouds and possibly dark matter and energy) calculated with direct grid methods can be prohibitively expensive. For example, the pseudospectral Fourier method for calculating the spatial derivatives requires a regular grid with a grid spacing of, say, $dx$ in the $x$ direction. We have seen above that according to the Nyquist theorem, $dx \le$ 750 km [equation (\ref{gs3})]. Assuming a 2D medium (shear, TM and gravitational waves do not propagate in 1D media) and $dx$ = $dz$ = 750 km, the grid to simulate the GW150914 event must have a minimum number of grid points $N_x \otimes N_z$ = 3.2 $\times$ 10$^{19}$ $\otimes$ 3.2 $\times$ 10$^{19}$ $\approx$ 10$^{39}$, which corresponds to 8 $\times$ 10$^{24}$ PB for a single field array in the computer code.

\section{Conclusions}

We use mathematical analogies to propose a theory for the propagation of gravitational waves mathematically equivalent to the theories of viscoelasticity and electromagnetism. One simplification is that we use a vector form of the tensor formulation of Einstein's equations, such as the
Maxwell's equations, for the limiting case of weak gravitational fields and moving sources with low velocities. The focus is on propagation, since the source is phenomenologically described by a modulated chirp function and this is a second simplification to establish the analogy.

We use the analogy to add a damping term to the gravitational equations, which corresponds to Ohm's law in electromagnetism and Maxwell's viscosity in viscoelasticity.
The physics is analyzed with a plane-wave analysis that defines the energy balance.
In general, the phase and energy velocity velocities are non-collinear, and the polarization of the cogravitational wave is linear and perpendicular to the plane defined by the propagation (wavenumber) and attenuation vectors, as cross-plane shear waves and TM electromagnetic waves (in isotropic media).
Transient wave fields from the source to the receiver are calculated with the Green function in homogeneous media and with a direct grid method in heterogeneous media. The grid method is successfully tested with the Green-function solution. These methods are feasible but require more powerful computers (possibly quantum computers) than those used today to solve realistic propagation events observed with current detectors.

\vspace{1cm}


\newpage

{\bf References}: 


\begin{verse} 

\item
Abbott,  B. P., Abbott, R., Abbott, T. D., Acernese, F., Ackley,  K., Adams,  C., et al., 2017, Observation of gravitational waves from a binary neutron star inspiral, Phys. Rev. Lett., 119 (16), 161101. 

\item
Ajith, P., Babak, S., Chen, Y., Hewitson, M., Krishnan, B., Sintes, A. M., Whelan, J. T., Br\"ugmann, B.,  Diener, P., Dorband, N., Gonzalez, J., Hannam, M., Husa, S., Pollney, D., Rezzolla, L., Santamar\'{\i}a, L., Sperhake, U., and Thornburg, J., 2008,  
Template bank for gravitational waveforms from coalescing binary black holes: Nonspinning binaries,
Phys. Rev. D 77, 104017; Erratum Phys. Rev. D 79, 129901 (2009).

\item
Asenjo, F. A., and Mahajan, S. M., 2020, Resonant interaction between dispersive gravitational waves and scalar massive particles, Phys. Rev. D, 101(6), 063010, doi:10.1103/PhysRevD.101.063010 

\item
Barnett, S. M., 2014,
Maxwellian theory of gravitational waves and their
mechanical properties,
New Journal of Physics, 16, 023027.

\item
Baym, G., Patil, S. P., and Pethick, C. J., 2017, 
Damping of gravitational waves by matter,
Phys. Rev. D, 96, 084033

\item
Behera, H., and Barik, N., 2017,
Attractive Heaviside-Maxwellian (vector) gravity from
special relativity and quantum field theory,  	
https://doi.org/10.48550/arXiv.1709.06876

\item
Behera, H., and Naik, P. C., 2004, Gravitomagnetic moments and
dynamics of Dirac's (spin 1/2) fermions in flat space-time
Maxwellian gravity, Int. J. Mod. Phys. A, 19, 4207--4229.

\item
Bland, D. R., 1960, The theory of linear viscoelasticity, Oxford, New York, Pergamon Press.

\item
Braginsky, V. B., Caves, C. M., Thorne, K. S., 1977, Laboratory experiments to test relativistic gravity, Phys. Rev. D., 15, 2047--2068.

\item
Buchen, P.W., 1971, Plane waves in linear viscoelastic media, Geophys. J. R. Astron. Soc. 23, 531--542.

\item
Ciubotariu, C. D., 1991, 
Absorption of gravitational waves, 
Physics Letters, Al58, 27--30.

\item
Carcione, J. M., 1996, Ground radar simulation for archaeological applications, Geo-
phys. Prosp., 44, 871--888.

\item
Carcione, J. M., 2022, Wave fields in real media. Theory and numerical simulation 
of wave propagation  in anisotropic, anelastic, porous and electromagnetic media, 4th edition, Elsevier.  

\item
Carcione, J. M., and Cavallini, F., 1995a, On the acoustic-electromagnetic analogy, Wave Motion, 21, 149--162.

\item
Carcione, J. M., Cavallini, F., 1995b, The generalized SH-wave equation, Geophysics 60, 549--555.

\item
Carcione, J. M., Liu, X., Greenhalgh, S., Botelho, M. A. B., and Ba, J., 2020, Quality factor of inhomogeneous plane waves, Phys. Acoust., 66, 598--603.

\item
Cattani, D. D., 1980, Linear equations for the gravitational
field, Nuovo Cimento B, Serie 11, 60B, 67--80.

\item
Daniels, D. J. (Ed.), 2004, Ground Penetrating Radar--2nd Edition.
Institution of Electrical Engineers, London, UK.

\item
Deidda, G. P., and Balia, R., 2001, 
An ultrashallow SH?wave seismic reflection experiment on a subsurface ground model
Authors, 66, 988--1328

\item
Einstein, A., and Rosen, N., 1937, On gravitational waves, J. of the Franklin Institute. 223(1), 43--54. 

\item
Forward, R. L., 1961, General relativity for the experimentalist, 
Proceedings of the Inst. Radio Eng., 49(5), 892--904.

\item
Gomes, H., and Rovelli, C., 2024, 
On the analogies between gravitational and electromagnetic radiative energy,
Philosophy of Physics, DOI: 10.31389/pop.58

\item
He, J.-H., 2021, GWsim: a code to simulate gravitational waves propagating in a potential well
Monthly Notices of the Royal Astronomical Society, 506(4), 5278--5293. 

\item
Heaviside, O., 1893, 
A gravitational and electromagnetic
analogy, 
The Electrician, 
Part I, 31, 281-282;
Part II, 31, 359.

\item
Hills, B. P., 2020, Gravito-electromagnetism and mass induction.

\item
Koene, E. M. F., 2020, 
A filtering approach to remove finite-difference
errors from wave equation simulations, 
PhD thesis. ETH Zurich.

\item
Krebes, E.S., Le, L.H.T., 1994, Inhomogeneous plane waves and cylindrical waves in anisotropic anelastic media, J. Geophys. Res. 99, 23899--23919.

\item
Peng, H., 1983, On calculation of magnetic type gravitation and experiments, Gen. Rev. Grav., 15(8), 725--735. 

\item
L\'opez, G., V., 2018, 
On Maxwell equations for gravitational field, 
J. Appl. Math. Phys., 6, 932--947. 

\item
Maxwell, J. C., 1865. A dynamical theory of the electromagnetic field, Philos. Trans. R. Soc.
Lond., 155, 459--512.

\item
McDonald, K. T., 1997, 
Answer to question \#49 (by D. Keeports, 1996, Am. J. Phys, 64, 1097). Why $c$ for gravitational waves?, Am. J. Phys. 65, 591--592.

\item
M\"uller, G., 1985, 
The reflectivity method: a tutorial, 
J. Geophys., 58, 153--174. 

\item
Pandey, S. S., Sarkar, A., Ali, A., and Majumdar, A. S., 2022,
Effect of inhomogeneities on the propagation of gravitational waves from binaries of compact objects,
Journal of Cosmology and Astroparticle Physics, DOI 10.1088/1475-7516/2022/06/021.

\item
Peatross, J. and Ware, M., 2024, 
Physics of light and optics,
Brigham Young University publications, 
\url{https://optics.byu.edu/docs/OpticsBook.pdf}

\item
Prather, C. B., 2020, 
Gravitational radiation: Maxwell-Heaviside formulation, 
Master thesis, 
University of Central Oklahoma.

\item
Price, R., Belcher, J., and Nichols, D., 2013, Comparison of electromagnetic and gravitational radiation: What we can learn about each from the other, Am J. Phys., 81, 575--584. 

\item
Roy, S., 2022, 
Nonorthogonal wavelet transformation for reconstructing gravitational wave signals, 
Phys. Rev. Res., 4, 033078 

\item
Scharpf, P., 2017, 
Simulation and visualization of gravitational waves
from binary black holes, 
Master thesis, University of Stuttgart.  

\item
Thid\'e, B., 2011, 
Electromagnetic field theory, 2nd edition, Dover. 

\item
Thorne, K., 1994, 
Gravitational waves, Proceedings of the Snow-
mass 95 Summer Study on Particle and Nuclear Astrophysics and
Cosmology, eds. E. W. Kolb and R. Peccei (World Scientific,
Singapore).

\item
Ummarino, G. A., and Gallerati, A., 2019,  
Exploiting weak field gravity-Maxwell symmetry in superconductive fluctuations regime, Symmetry, 11, 1341. 

\item
Wald, R. M., 1984, General relativity. University of Chicago Press.

\item
Weber, J., 1961, General relativity and gravitational waves, New York: Interscience.

\item
Weinberg, S., 2004, 
Damping of tensor modes in cosmology, 
Phys. Rev. D., 69, 023503. 

\end{verse}


\newpage

\vspace{2cm}

\begin{center}
{\bf Table 1. Viscoelastic-electromagnetic-gravitational analogy}
\end{center}
\[
\begin{tabular}{| l l l|}
\hline
viscoelastic (VE) & electromagnetic (EM) & gravitational (GR) \\
\hline
$(-\sigma_{23}, 0, \sigma_{12})^\top$ = stress &{\bf E} = electric field & {\bf g} = gravitational field \\
$(0, v, 0)^\top$ = particle velocity & {\bf H} = magnetic field & {\bf h} = cogravitational field \\
$(0, \rho v, 0)^\top$ = momentum density &{\bf B} = magnetic induction & {\bf b } = cogravitational induction \\
$(0, f_2, 0)^\top$ = elastic source & {\bf J} = electric source & {\bf m} = gravitational source \\
$R^{-1}$ = compliance & $\epsilon$ = permittivity & $a$ = $1/(4 \pi G)$ \\
$\rho$ = mass density & $\mu$ = magnetic permeability & $b$ = $4 \pi G/c^2$ \\
$\eta^{-1}$ = fluidity & $\sigma$ = conductivity & $d$ = gravitational damping \\
$\rho$ = mass density  & $\rho_e$ = charge density & $\rho$ = mass density \\
 \hline
\end{tabular}
\]

\begin{singlespace}

\begin{center}
{\bf Table 2. Symbols in SI units.}
\end{center}

\hspace{2cm} ----------- Viscoelastic (VE)

\hspace{2cm} $(-\sigma_{23}, 0, \sigma_{12})^\top$  [kg/m/s$^2$]

\hspace{2cm} $(0, v, 0)^\top$ [m/s]

\hspace{2cm} $(0, \rho v, 0)^\top$  [kg/s/m$^2$]

\hspace{2cm} $(0, f_2, 0)^\top$ [kg/s$^2$/m$^2$]

\hspace{2cm} $R^{-1}$ [m s$^2$/kg]

\hspace{2cm}  $\rho$ [kg/m$^3$]

\hspace{2cm}  $\eta^{-1}$ [m s/kg]

\hspace{2cm}  $c = \sqrt{R/\rho}$ = high-frequency velocity [m/s] 

\hspace{2cm}  Energy densities [J/m$^3$] = [kg/m/s$^2$] 

\hspace{2cm}  Umov-Poynting vector [kg/s$^3$] 

\hspace{2cm} ----------- electromagnetic (EM)

\hspace{2cm} ${\bf E}$ [V/m] = [kg m/s$^3$/A]

\hspace{2cm} ${\bf H}$ [A/m]

\hspace{2cm} ${\bf B} = \mu {\bf H}$ [kg/s$^2$/A]

\hspace{2cm} ${\bf J}$ [A/m$^2$]

\hspace{2cm} $\epsilon$ in units of $\epsilon_0$ = 8.854 $\times$ 10$^{-12}[$C/(V m)] = [s$^4$ A$^2$/kg/m$^3$]

\hspace{2cm}  $\mu$ in units of $\mu_0$ = 4 $\pi$ 10$^{-7}$  [H/m] = [kg m/s$^2$/A$^2$]

\hspace{2cm}  $\sigma$ [S/m] = [s$^3$ A$^2$/kg/m$^3$]

\hspace{2cm}  $c = (\mu \epsilon)^{-1/2}$ = optical (high-frequency) velocity [m/s] 

\hspace{2.25cm} = 2.99792458 $\times$ 10$^8$ m/s $\approx$ 30 cm/ns

\hspace{2cm}  $\rho_e$ [C m$^{-3}$] = [A s m$^{-3}$] 

\hspace{2cm} ----------- gravitational (GR)

\hspace{2cm} ${\bf g}$ [m/s$^2$]

\hspace{2cm} ${\bf h}$ [kg/m/s]

\hspace{2cm} ${\bf b}$ = $b {\bf h}$  [1/s]

\hspace{2cm} ${\bf m}$ [kg/m$^2$/s]

\hspace{2cm} $a$ [s$^2$ kg/m$^3$]

\hspace{2cm}  $b$ [m/kg]

\hspace{2cm}  $d$ [s kg/m$^3$]

\hspace{2cm}  $c = (a b)^{-1/2}$ = high-frequency velocity [m/s] 

\hspace{2.25cm} = 2.99792458 $\times$ 10$^8$ m/s $\approx$ 30 cm/ns

\hspace{2cm}  $G$ = gravitational constant = 6.6743 10$^{-11}$ [N m$^2$/kg$^2$ = m$^3$/s$^2$/kg]

\hspace{2cm}  $\rho$ [kg m$^{-3}]$ 

\end{singlespace}
                            
\newpage

\appendix

\section{Source wavelet}

Growing and decaying chirp signals can approximate parts of gravitational source time histories (Roy, 2022). 
For a linear chirp
\begin{equation} \label{A1}
h (t) = \sin \left[\phi_0 +2 \pi \left( \frac{1}{2} \nu t^2 + f_0 t \right) \right] , 
\end{equation}
where $\phi_0$ is the initial phase, $t_0$ is the initial time, 
\begin{equation} \label{A2}
\nu = (f_1 - f_0) /T,  
\end{equation}
is the chirp rate, 
$f_ 1$ is the final frequency and $T = t_1-t_0$ is the time it takes to sweep from $f_0$ to $f_1$.

On the other hand, the signal of an exponential chirp is 
\begin{equation} \label{A3}
h (t) = \sin \left[\phi_0 +2 \pi f_0 \left( \frac{\exp[\zeta (t-t_0) ]-1}{\zeta} \right) \right] , 
\end{equation}
where
\begin{equation} \label{A4}
\zeta = \frac{\ln(f_1/f_0)}{t_1-t_0} , 
\end{equation}
If $f_1 = f_0$, $\zeta$ = 0 and the argument of the sine is $\phi_0 + 2 \pi f_0 (t-t_0)$. 

An hyperbolic chirp has the equation 
\begin{equation} \label{A5}
h (t) = \sin \left[\phi_0 +2 \pi f_0 \tau \ln \left( 1 - \frac{t}{\tau} \right)  \right] , 
\end{equation}
where
\begin{equation} \label{A6}
\tau = \frac{f_1 (t-t_0)}{f_1-f_0} . 
\end{equation}

Morlet wavelets are also employed to characterize gravitational signals (Roy, 2022). A typical equation is 
\begin{equation} \label{A7}
h (t) =  \cos ( 2 \pi f_c (t-t_0) ] \exp[- a_M (t-t_0)^2), 
\end{equation}
where $f_c$ is the central frequency. 

A source time history of a coalescence of two binary black holes is composed of inspiral, merger and ring down. 
If ring down starts at $t=t_2$, we apply to $h(t)$ growth and damping factors as follows 
\begin{equation} \label{A8i}
\begin{array}{l}
\exp[a_1 (t-t_0)], \ \ \  a_1 > 0, \ \ \  t<t_2 , \\ \\
\exp[a_2 (t-t_2)], \ \ \ a_2 < 0, \ \ \  t \ge t_2 .
\end{array}
\end{equation}
We then have the source signal
\begin{equation} \label{A8}
\begin{array}{ll}
S(t) & = h (t) \exp[a_1 (t-t_0)], \ \ \  t<t_2 , \ \ \ \mbox{inspiral and merger} 
\\ \\
 & = h (t) \exp[a_2 (t-t_2)] \exp[a_1 ( t_2 - t_0)], \ \ \  t \ge t_2 , \ \ \ \mbox{ring down} 
\end{array}
\end{equation}
where $h$ corresponds to the above chirp time functions. The second exponential in the ring down guarantees that $S$ is continuous at $t_2$. 

\section{Green's function solution}

Frequency-domain solutions corresponding to equation (\ref{p7}) in a homogeneous 
medium can easily be obtained. Consider the lossless gravitational equations (\ref{p22}) ($d$ = 0) with a source $m_2$
\begin{equation} \label{B1}
\begin{array}{ll}       
\partial_1 g_3 - \partial_3 g_1 = b \partial_t h + m_2 ,  \\  \\
- \partial_3 h = a \partial_t g_1 ,   \\   \\
\partial_1 h =  a \partial_t g_3  , 
\end{array}
\end{equation}
Eliminating the gravitational components we have 
\begin{equation} \label{B1}
\frac{1}{a} \Delta h - b \partial_{tt} h = \partial_t m_2, 
\end{equation}
Transforming the equation to the frequency domain and considering the Green function $G$ gives 
\begin{equation} \label{B2}
\frac{1}{a} \Delta G  + b \omega^2 G = - 8 \delta (x-x_0) \delta (z-z_0) 
\end{equation}
or
\begin{equation} \label{B3}
\Delta G  + k^2 G = -8 a\delta (x-x_0) \delta (z-z_0), \ \ \ k = \myfrac{\omega}{c}, \ \ \ c = \myfrac{1}{\sqrt{ab}} . 
\end{equation}
The constant $-8$ is introduced for convenience.
The solution to (\ref{B3}) is 
\begin{equation} \label{B2}
G (r,r_0, \omega, k) = - 2 \ii a H_0^{(2)} \left( k r \right)  , 
\end{equation}
where $H_0^{(2)}$ is the zero-order Hankel function of the second kind 
(e.g. Koene, 2020, Eq. 2.76; Carcione, 2022, Eqs. 7.561 and 7.563; Koene, 2020), $r_0 = (x_0, z_0,)$ is the source location, and 
\begin{equation} \label{B3}
r = \sqrt{ (x - x_0)^2 + (z - z_0)^2 }.
\end{equation}
The solution in lossy media is obtained by invoking the correspondence principle (Bland, 1960), 
i.e., by substituting the real $a$ with the complex $\bar a$ given in equation (\ref{p6}).
We set $G ( - \omega) = G^\ast ( \omega)$, where the superscript
``$\ast$" denotes complex conjugation. 
This equation ensures that the inverse Fourier 
transform of the Green's function is real.
The frequency-domain solution is then given by 
\begin{equation} \label{B4}
h (\omega) = \ii \omega m_0 G (\omega ) S (\omega ) ,
\end{equation}
where $S$ is the Fourier transform of the source time history, $m_0$ is the strength of the source and the factor $\ii \omega$ is due to
the first-order time derivative of the source in equation (\ref{B1}). 
Because the Hankel function has a singularity at 
$\omega$ = 0, we assume
$G=0$ for $\omega$ = 0, an approximation that does not have a significant effect on the 
solution (note, moreover, that 
$S(0)$ $\approx$ 0). The time-domain solution 
$ h(t)$ is obtained by a discrete inverse Fourier transform. 

On the other hand, in the lossless case ($d$ = 0), the solution can be computed with a numerical temporal convolution between the source time history $S(t)$ and the time-domain Green function as 
\begin{equation} \label{B5}
h (t) = G (t ) \ast S (t) , 
\end{equation}
where
\begin{equation} \label{B6}
G (r,r_0, t) = \myfrac{4 a}{\pi}\left( t^2 - \frac{r^2}{c^2} \right)^{-1/2} H \left( t - \frac{r}{c} \right) 
\end{equation}
(Koene, 2020, Eq. 2.79; Carcione, 2022, Eq. 3.202), where $H$ is the Heaviside function. In the lossy case, the convolution is not possible because there is no analytical solution for the Green function in the time domain. 

\newpage

\begin{figure}
\vspace{-2cm}
\includegraphics[scale=0.5]{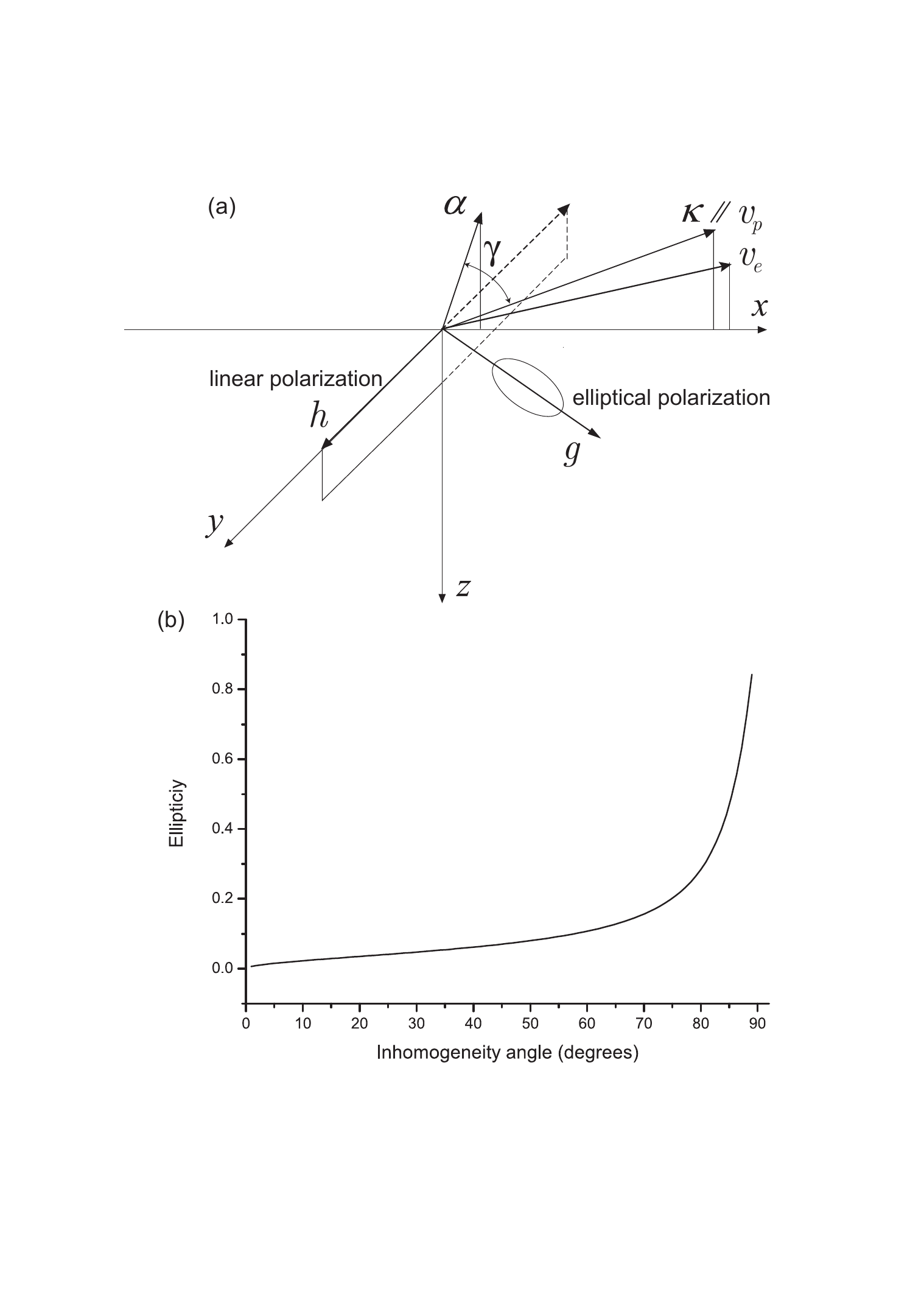}
\vspace{-3cm}
\caption{Inhomogeneous plane wave (a). Vectors ${\BM \kappa}$, ${\BM \alpha}$ and ${\bf v}_e$ are coplanar and 
${\BM \kappa}$ and ${\bf v}_p$ are collinear (all these vectors lie in the $(x,z)$-plane). Vector ${\bf h}$ (polarization) is perpendicular to the  $(x,z)$-plane. For homogeneous plane waves ($\gamma$ = 0), ${\BM \kappa}$, ${\BM \alpha}$, ${\bf v}_e$ and ${\bf v}_p$ are collinear and ${\bf v}_e$ = ${\bf v}_p$. Ellipticity of the ${\bf g}$ vector lying in the $(x,z)$-plane (b). The analogy holds with SH and TM waves.
}
\end{figure}

\begin{figure}
\vspace{-3.cm}
\includegraphics[scale=0.4]{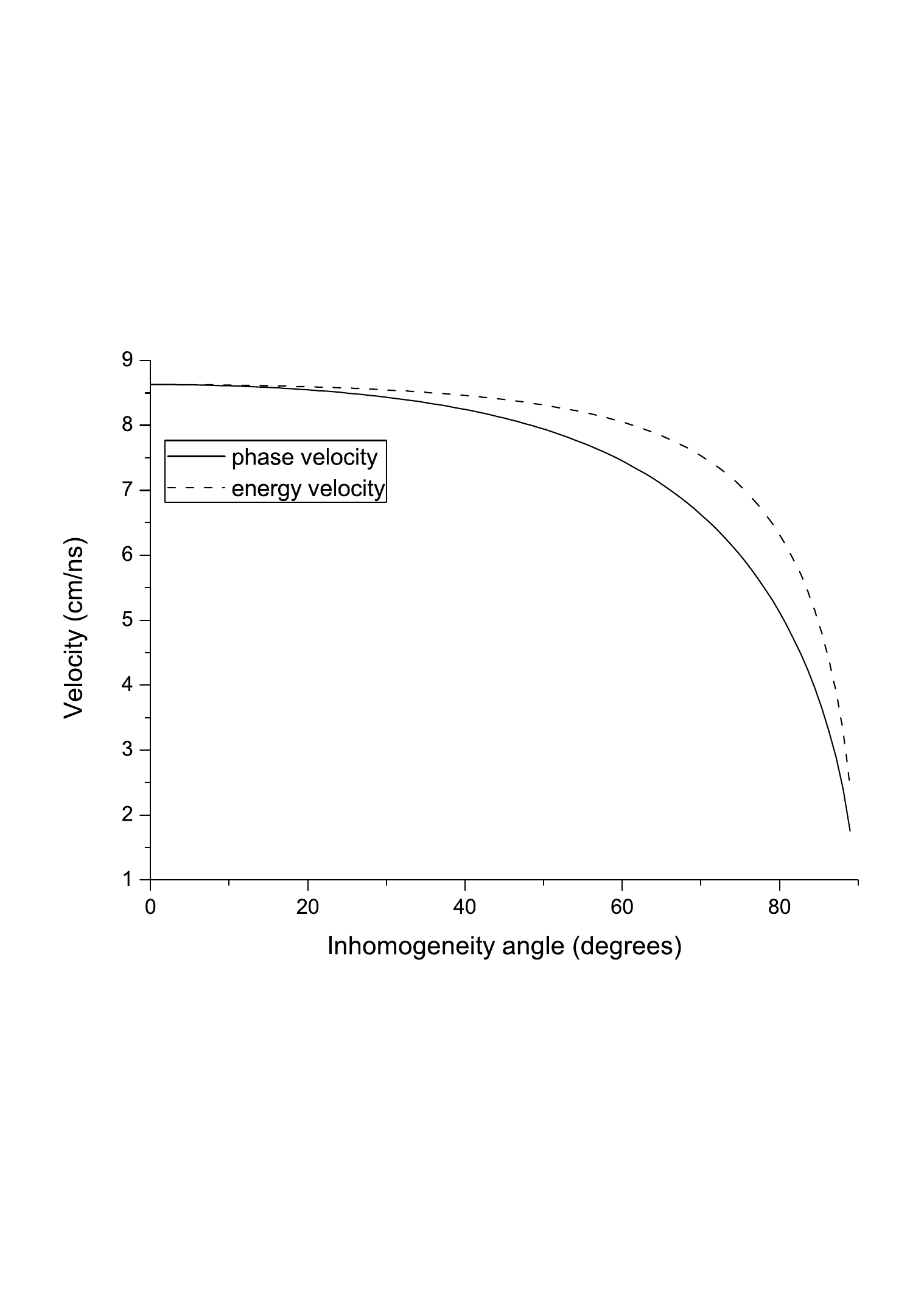}
\vspace{-3.5cm}
\caption{EM phase and energy velocities $v_p$ and $v_e$ in wet clay as a function of the inhomogeneity angle $\gamma$.  
}
\end{figure}

\begin{figure}
\vspace{0.cm}
\includegraphics[scale=0.45]{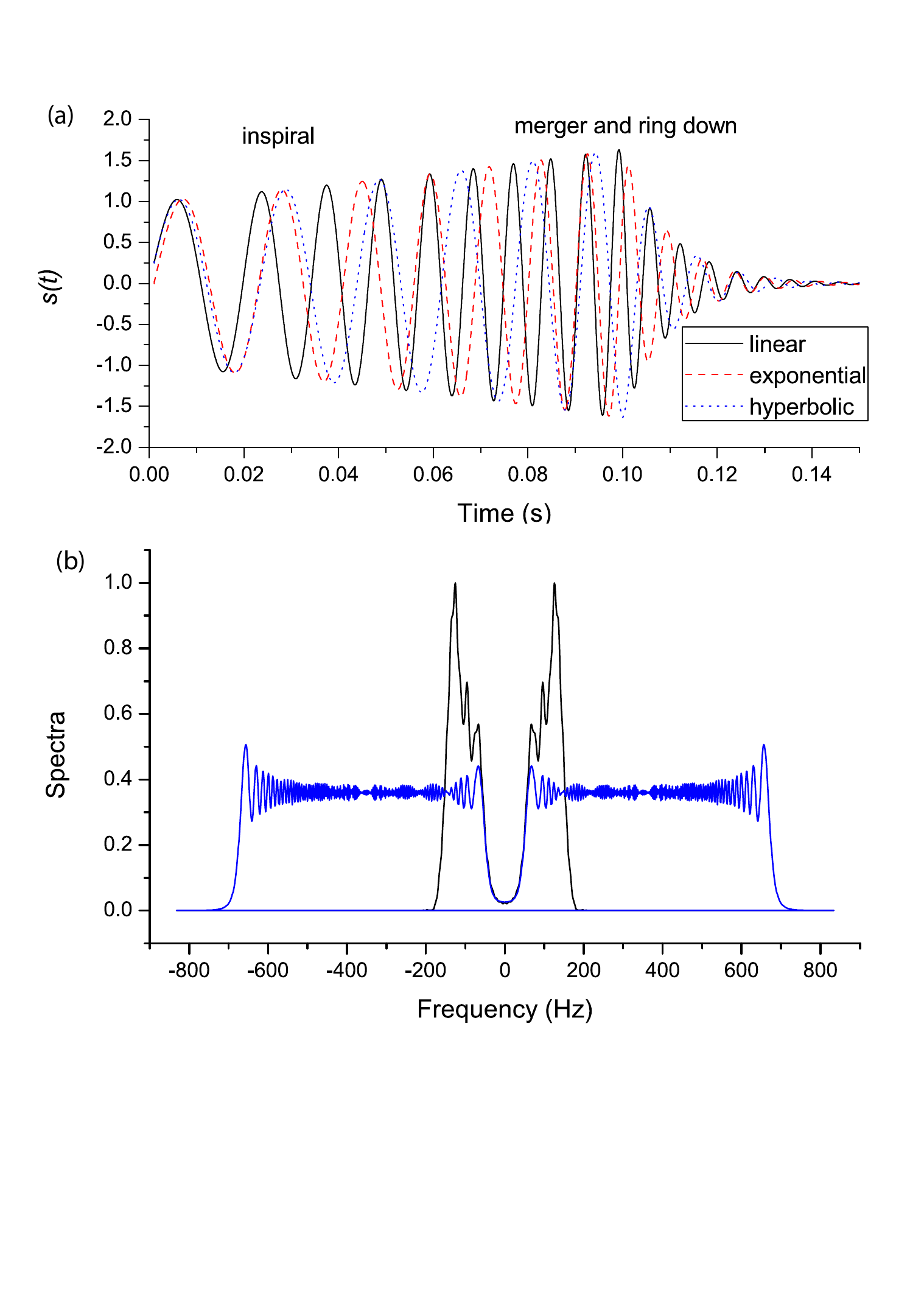}
\vspace{-3.0cm}
\caption{Heuristic gravitational source time history (equation (\ref{A8})) corresponding to different chirps signals (a) and normalized spectrum for the linear chirp (b). The blue line corresponds to $a_1 = a_2 $ = 0 (no growth and damping).}
\end{figure}

\begin{figure}
\vspace{0.cm}
\includegraphics[scale=0.45]{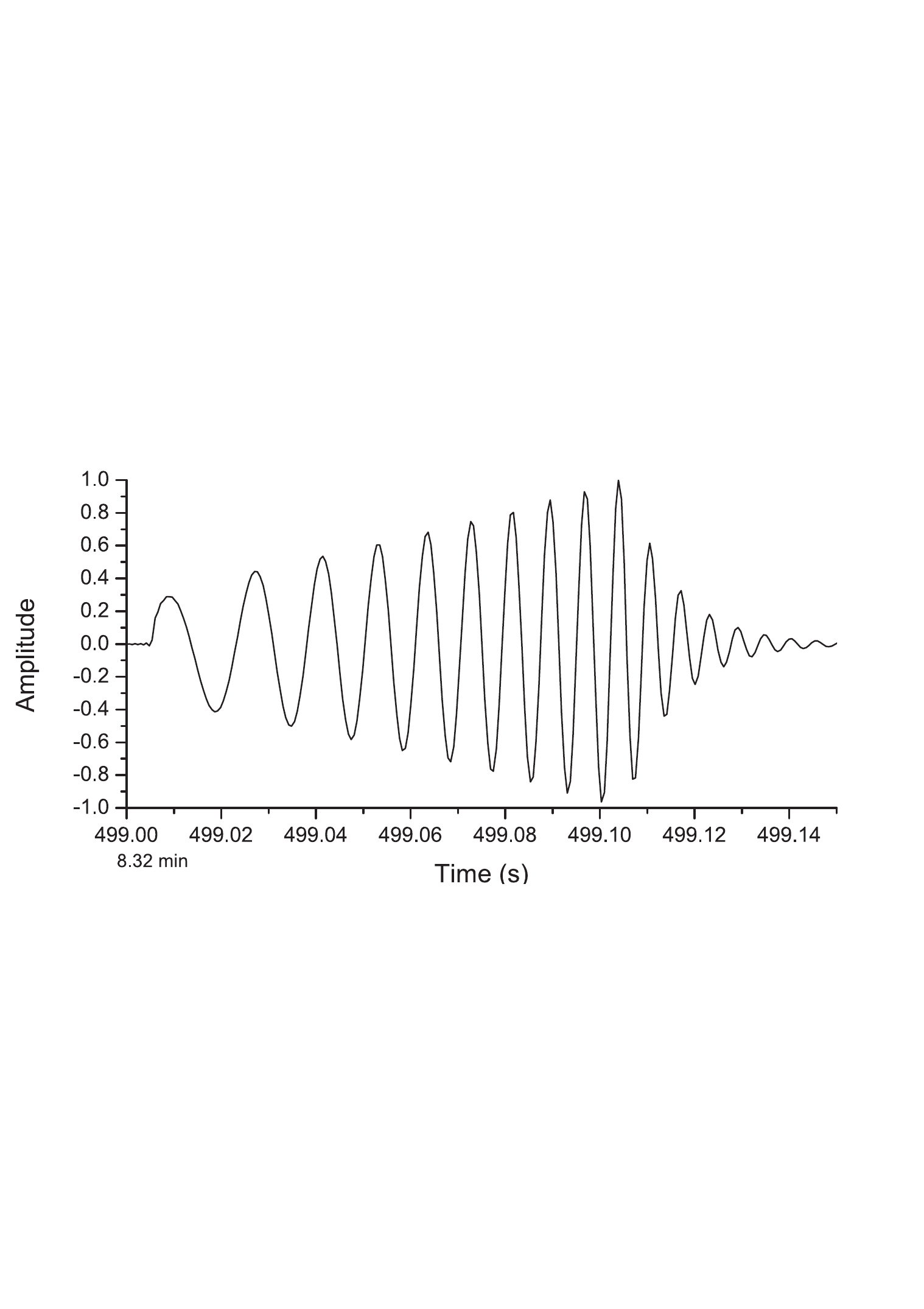}
\vspace{-4.5cm}
\caption{Solution of component $h$ using the Green functions (\ref{B2}) with $d$ = 0 (no loss). The signal has traveled the Sun-Earth distance at the velocity of light. The field is normalized.}
\end{figure}

\begin{figure}
\vspace{0.cm}
\includegraphics[scale=0.4]{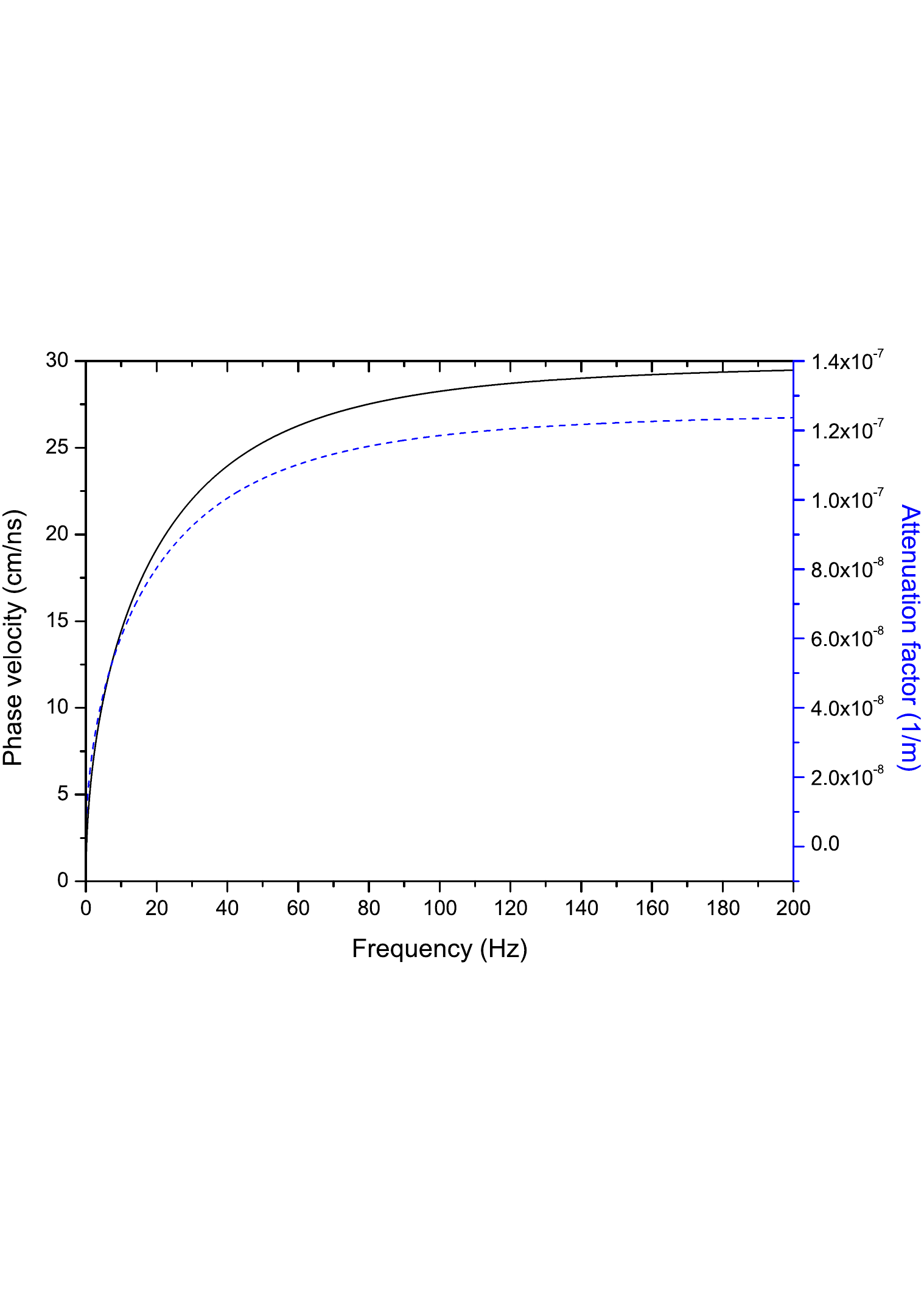}
\vspace{-3.5cm}
\caption{Phase velocity (equation \ref{p11})) (solid line) and attenuation factor (equation \ref{p12})) (dashed line) for homogeneous waves as a function of frequency.}
\end{figure}

\begin{figure}
\vspace{0.cm}
\includegraphics[scale=0.45]{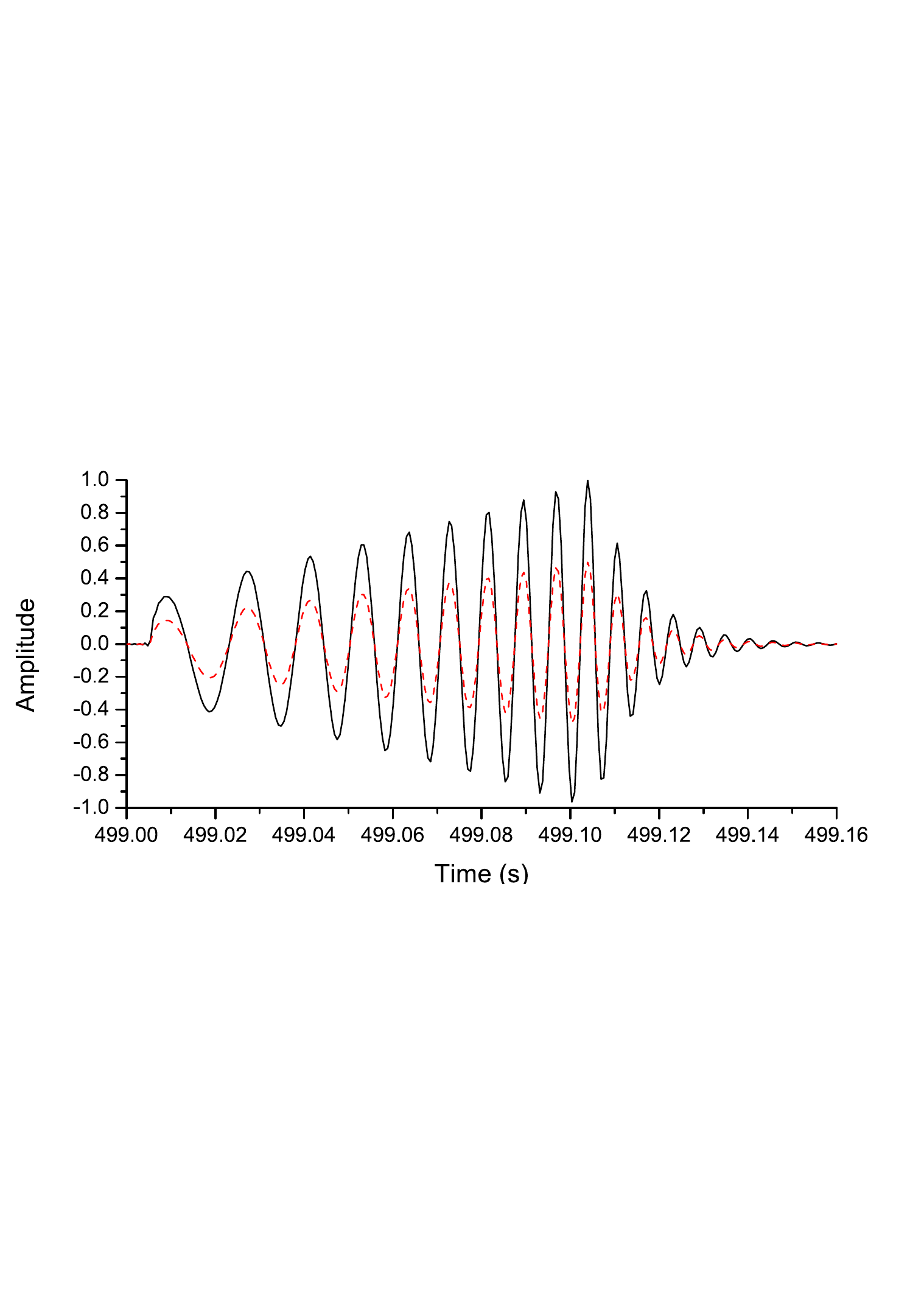}
\vspace{-4.9cm}
\caption{Signal shown in Figure 4 (solid line) and signal with attenuation ($Q$ = 3 $\times$ 10$^5$) (dashed red line).}
\end{figure}

\begin{figure}
\vspace{-2.cm}
\includegraphics[scale=0.45]{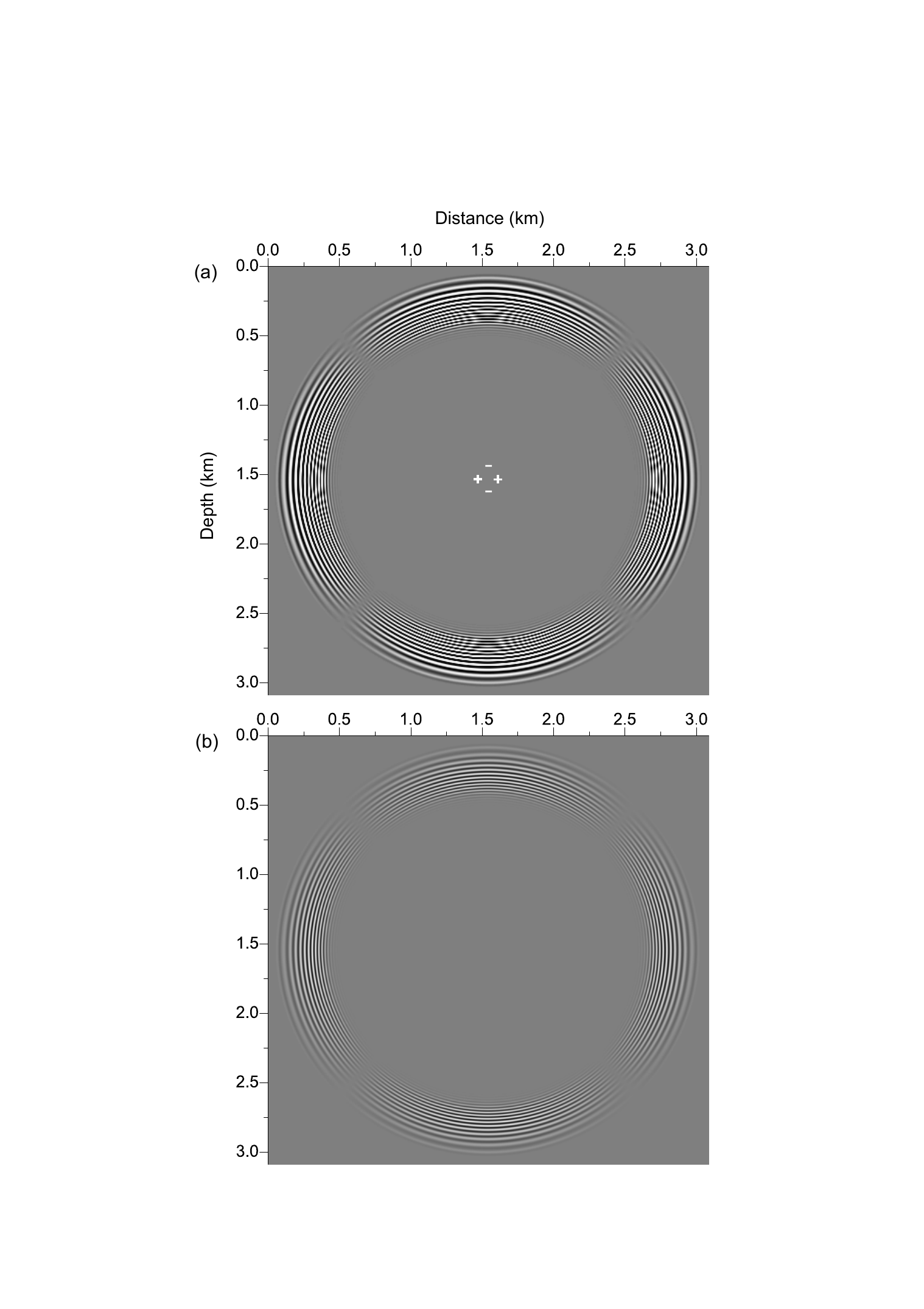}
\vspace{-1.5cm}
\caption{Snapshots of the component $v$ in the lossless elastic (a) and lossy viscoelastic (b) cases, due to a quadrupole source.}
\end{figure}

\begin{figure}
\vspace{-2.cm}
\includegraphics[scale=0.45]{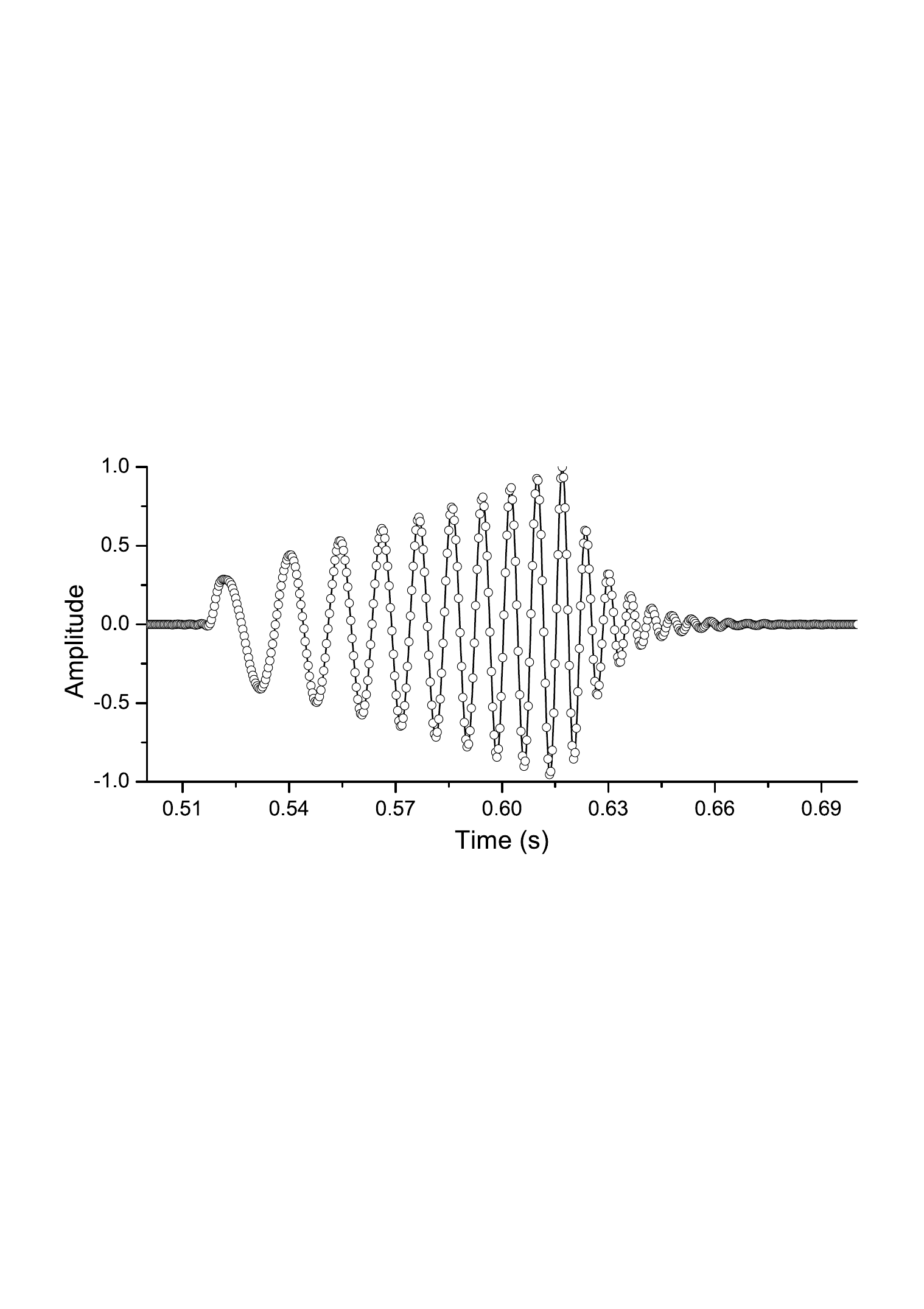}
\vspace{-5.4cm}
\caption{Test of the grid-method solution (open circles) with the Green-function solution (solid line). We compare the $h$ field at a distance of 0.52 light-seconds from a point source.}
\end{figure}

\begin{figure}
\vspace{-1.cm}
\includegraphics[scale=0.4]{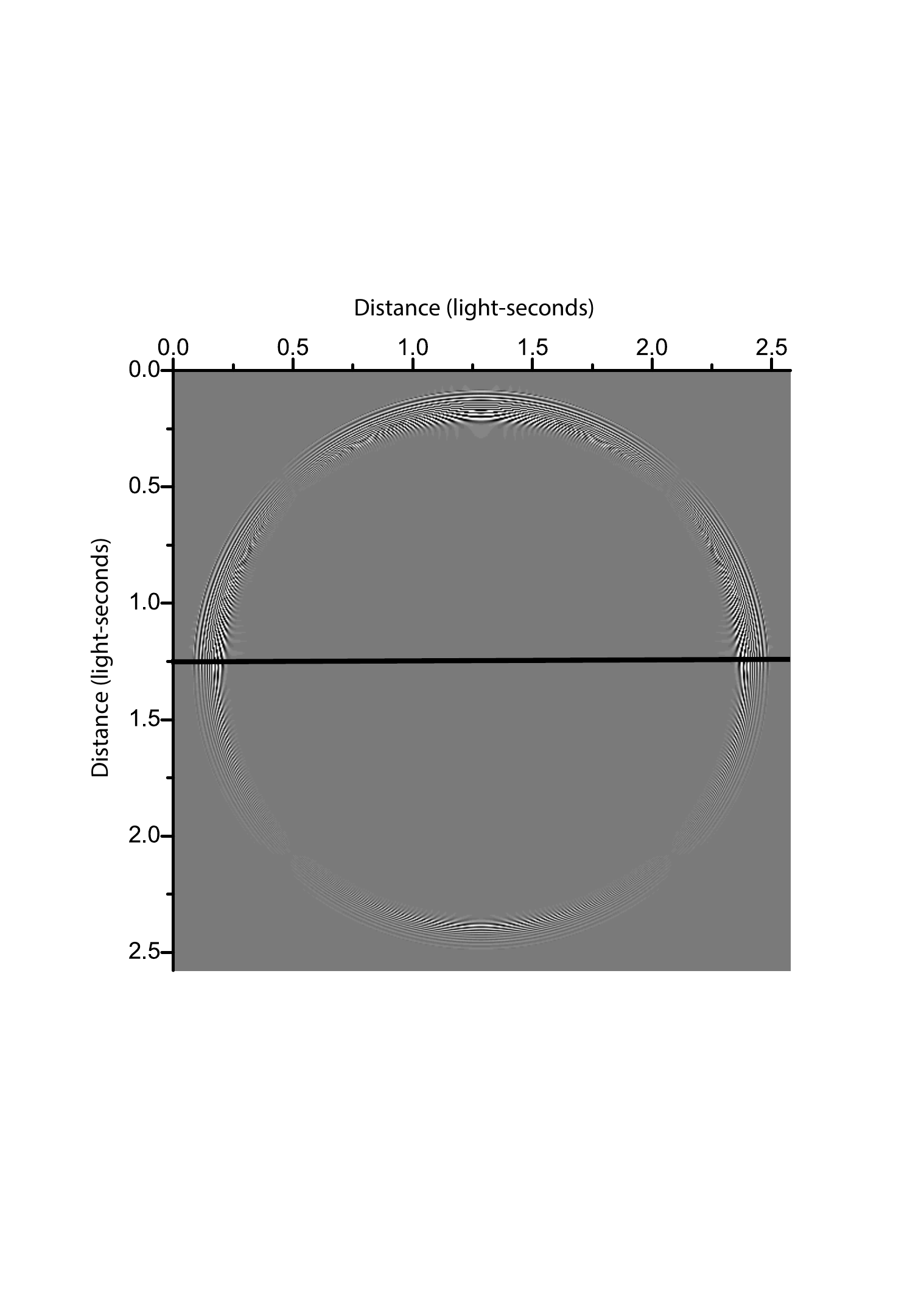}
\vspace{-3.5cm}
\caption{Snapshot of the component $h$ of a gravitational wave due to a quadrupole source. Above the interface there is no attenuation ($d$ = 0) and below the interface $Q$ = 150 ($d$ = 4 $\times$ 10$^9$ s kg/m$^3$).}
\end{figure}

\begin{figure}
\vspace{-1.cm}
\includegraphics[scale=0.4]{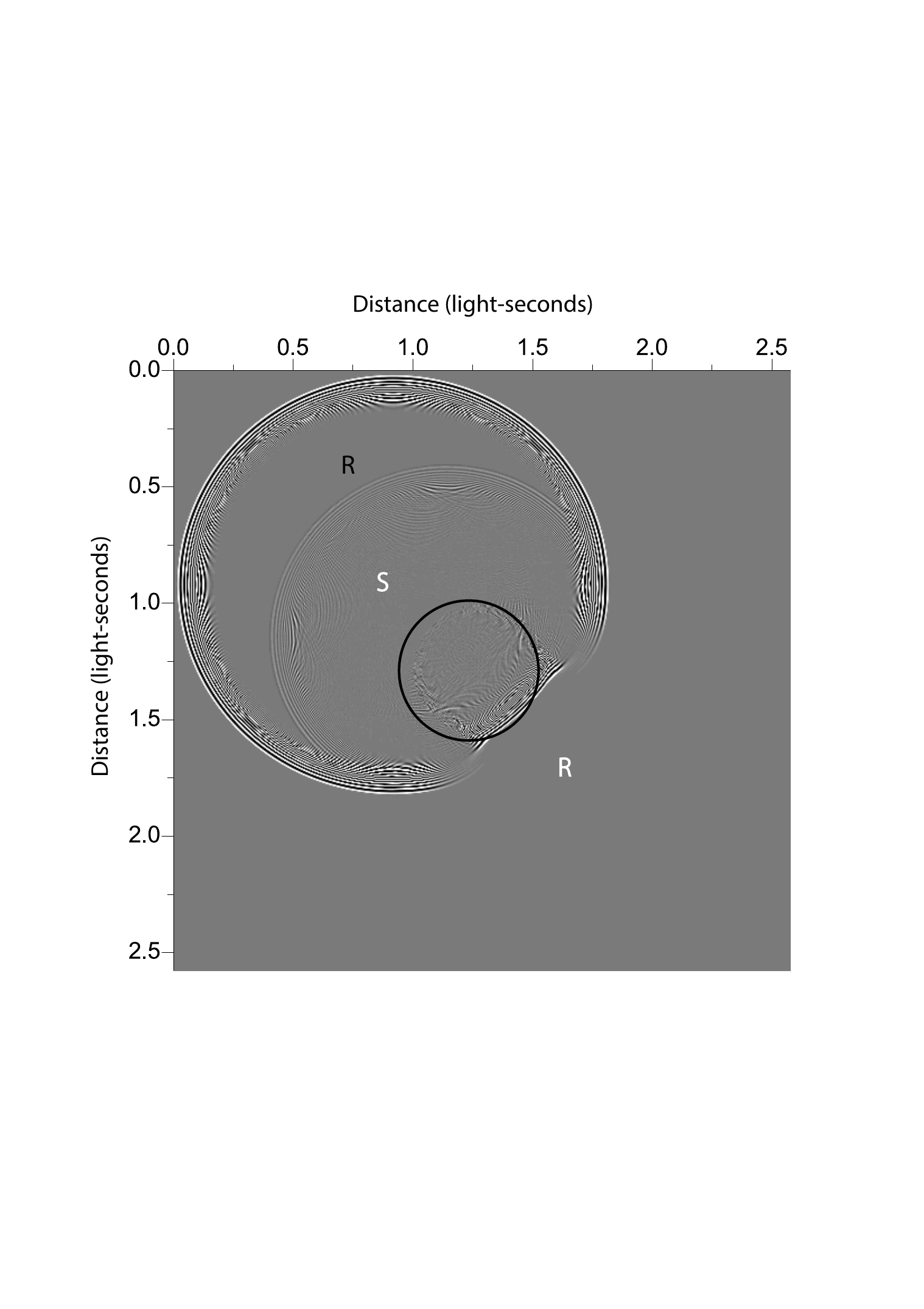}
\vspace{-3.5cm}
\caption{Snapshot at 0.9 s of the component $h$ of a gravitational wave due to a point source ($d$ = 0). The velocity within the circle is 0.8 $c$ and the source (white S) and receiver (white R) are indicated. Black R denotes the reflected event from the circle. 
}
\end{figure}

\begin{figure}
\vspace{-3.cm}
\includegraphics[scale=0.5]{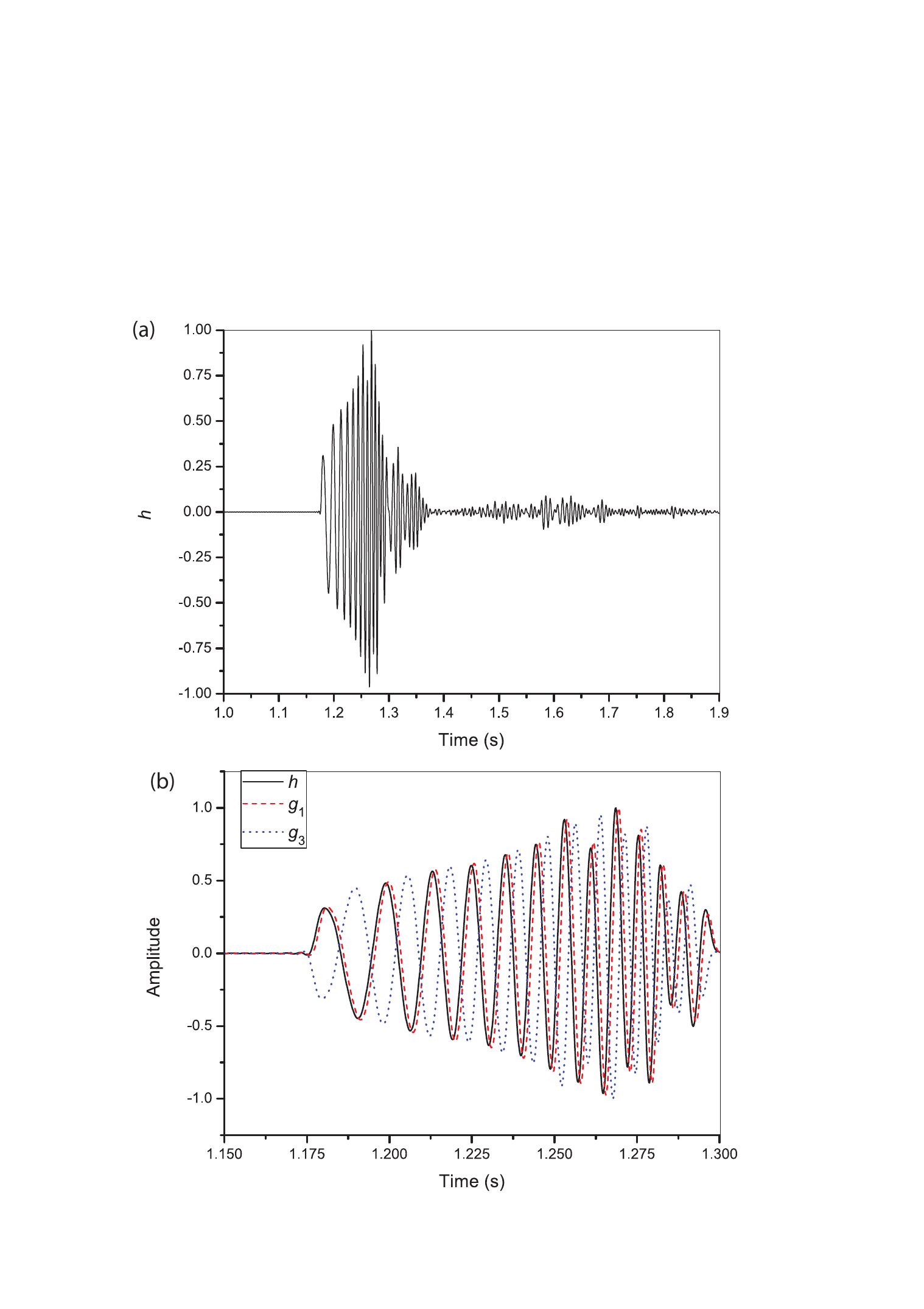}
\vspace{-1.5cm}
\caption{Time histories of the component $h$ (a) and three components (b) of a gravitational wave due to a point source ($d$ = 0). 
The fields are normalized, and the amplitude relation in (b) is $h/g_1/g_3$ = 1/ 2.6 $\times$ 10$^{-18}$ / 2.7 $\times$ 10$^{-18}$ (in kg s/m$^2$).
}
\end{figure}

\end{document}